\documentclass[10pt]{revtex4-1}
\usepackage{amssymb}
\usepackage{amsmath}
\usepackage{array} % for figures!
\usepackage{hyperref,bm,color}
\usepackage{graphicx}
\usepackage{multirow}
\begin{document}

\title[An $O(N^{3})$ implementation of Hedin's $GW$]{
An $O(N^{3})$ implementation of Hedin's $GW$ approximation for molecules} 

\author{D.~Foerster}%
\email{d.foerster@cpmoh.u-bordeaux1.fr}
\affiliation{CPMOH/LOMA, Universit\'e de Bordeaux 1, 351 Cours de la Liberation, 33405 Talence, France}

\author{P.~Koval}%
\email{koval.peter@gmail.com}

\author{D. S\'anchez-Portal}%
\email{sqbsapod@sq.ehu.es}

\affiliation{Centro de F\'{\i}sica de Materiales CFM-MPC, 
Centro Mixto CSIC-UPV/EHU, Paseo Manuel de Lardizabal 5, E-20018 San Sebasti\'an, Spain}
\affiliation{Donostia International Physics Center (DIPC), 
Paseo Manuel de Lardizabal 4, E-20018 San Sebasti\'an, Spain}



% \author{D.~Foerster\\
% CPMOH/LOMA, Universit\'e de Bordeaux 1, \\
% 351 Cours de la Liberation, 33405 Talence, France \\
% \\
% P.~Koval and D. S\'anchez-Portal \\
% Centro de F\'{\i}sica de Materiales CFM-MPC,
% Centro Mixto CSIC-UPV/EHU, \\ and Donostia International
% Physics Center (DIPC), Paseo Manuel de Lardizabal 5, \\
% E-20018 San Sebasti\'an, Spain }

%\maketitle

%% uncomment this for revtex
\begin{abstract}
We describe an implementation of Hedin's $GW$ approximation for
molecules and clusters, the complexity of which scales as $O(N^3)$
with the number of atoms. Our method is guided by two strategies: 
\textit{i}) to respect the locality of the underlying electronic interactions
and  \textit{ii}) to avoid the singularities of Green's functions by
manipulating, instead, their spectral functions using fast Fourier transform
methods. To take into account the locality of the electronic interactions,
we use a local basis of atomic orbitals and, also, a local basis in
the space of their products. 
We further compress the screened Coulomb interaction into a space of lower
dimensions for speed and to reduce memory requirements.
The improved scaling of our method with respect to most of
the published methodologies should facilitate $GW$ calculations
for large systems. Our implementation is intended as a step forward
towards the goal of predicting, prior to their synthesis,
the ionization energies and electron affinities of the large
molecules that serve as constituents of organic semiconductors.
\end{abstract}

\maketitle

\section{Introduction}

Lars Hedin's $GW$ method~\cite{Hedin} is an approximate 
treatment of the  propagation of electrons in condensed matter
where an electron interacts with itself via a Coulomb 
interaction that is screened by virtual
electron-hole pairs. 
In periodic semiconductors, the $GW$ approximation is known to 
lead to surprisingly accurate gaps \cite{Schilfgaarde-Kotani-Faleev:2006},
while for finite clusters and molecules it provides 
qualitatively correct values of ionization
energies and electron affinities \cite{LouieRohlfing}. 
Hedin's $GW$ approximation is also needed, as a first step, 
when using the Bethe-Salpeter equation to find the optical properties
of systems in which the Coulomb interaction is only weakly screened. 

The present work is motivated by the rapid progress, during 
the last decade, in the field of organic semiconductors, especially 
in organic photovoltaics
and organic luminescent diodes~\cite{OrganicsReview}. 
To optimize such systems, it would be useful to know the key 
properties of their molecular constituents
before actually synthesizing them. 
In order to make such predictions it is necessary to develop algorithms with 
a favorable complexity scaling, since many of the technologically relevant
molecules are fairly large.
The method presented here is a step forward along this direction.
Its $O(N^3)$ scaling, with $N$ the number of atoms in the molecule, 
is an improvement over most existing methodologies.

While computational techniques for treating the $GW$ 
approximation for clusters and molecules have become 
sophisticated enough for treating molecules of interest in 
photovoltaics  \cite{Blase} or in the physiology of vision
\cite{RohlfingRetinol}, such calculations remain computationally expensive. 
The scaling with the number of atoms of these recent calculations
has not been published. However, in many cases  it is unlikely to
be better than  $O(N^4)$~\cite{private-comm-scaling}.
A recently published method for computing  total energies of molecules that uses
the random phase approximation also has $O(N^4)$ scaling~\cite{Furche}.
Actually, at this point it is difficult for us to envisage
a scaling exponent less than three because the construction
of the screened Coulomb interaction ---
the central element of the $GW$ approach --- requires
inverting a matrix of size $O(N)$ which, in general,
takes $O(N^3)$ operations.

The algorithm described in this paper
is based on two main ingredients:
\textit{i}) respecting the locality of the underlying
interactions and, \textit{ii}) the use of
spectral functions to describe the frequency/time
dependence of the correlators. The latter ingredient allows
for the use of the fast Fourier transform (FFT) to accelerate
the calculations, while the former idea of respecting locality
has been also at the heart of other efficient $GW$ methods, like
the successful ``space-time approach'' for periodic systems~\cite{Godby}.

Our method is based upon the use of spatially localized basis 
sets to describe the electronic states within
the linear combination of atomic orbitals (LCAO) technique.
In particular, we have implemented our method as a post processing
tool of the SIESTA code~\cite{siesta}, although 
interfaces with other LCAO codes should be simple to construct. 
The precision of the LCAO
approach is difficult to control and to improve, 
but a basis of atom centered local orbitals is useful for systems 
that are too large to be
treated by plane-wave methods~\cite{reviewPayne}. 
In order to solve Hedin's equations we construct a basis set
that gets rid of the over completeness of the orbital products
while keeping locality.
In molecular computations this is frequently done
through a fitting procedure (using Gaussians or
other localized functions). We use an alternative
mathematical procedure~\cite{DF,DF+PK} that dispenses
with this fitting  and defines a basis of \textit{dominant products}.
The basis of dominant products was
instrumental to develop an efficient linear response code 
for molecular absorption spectra \cite{PK+DF+OC,Licence}.
In the present paper we have developed 
an additional, non local
compression technique that further reduces
the size of the product basis. The compression allows to store the whole
matrix representation of the screened Coulomb interaction at all times/frequencies
and needs much less memory. 
Moreover, the compression strongly accelerates the 
calculation of the screened Coulomb interaction 
because it involves a matrix inversion. This leads
to a gain in computational efficiency 
which is even more important than that associated 
with the reduction of the needed memory.

Of course there are other methods that use a localized basis different
from LCAO and, thus, equally appropriate for dealing with clusters and
molecules while exploiting locality. One method uses a lattice in real
space~\cite{RealSpaceChelikowsky}. Another method uses wavelets that
represent a useful compromise between localized and extended
states~\cite{BigDFT}. Localized Wannier orbitals obtained from 
transforming plane waves~\cite{Wannier} have also been used  in
$GW$ calculations~\cite{Baroni$GW$,DanishWannierMolecules}.
In this paper we use a basis of
dominant functions to span the space of products of atomic orbitals \cite{DF}
and we use a compression scheme to deal with
the screened interaction.
It is clear, however, that some of the ideas and techniques of the present 
paper can be combined with the alternative approaches quoted above. 

The actual implementation of the algorithm that we report in this
paper can be considered as a ``proof of principle'' only and
the prefactor of our implementation leaves room for further improvement.
Therefore, we validate our method with molecules of moderate size
(we consider molecules of only up to three aromatic rings: benzene,
naphthalene and anthracene),  leaving further improvements and
applications to molecules of larger sizes for a future publication.

This paper is organized as follows.
In section \ref{s:elementary-aspects-$GW$}
we recall the equations of the $GW$ approximation. 
In section \ref{s:tensor-form} we rewrite the $GW$ approximation
for molecules in tensorial form.
Section \ref{s:instant-scr-inter} describes the instantaneous component of the self-energy, 
while in section \ref{s:sf-techniques} 
we describe a spectral function technique for solving these tensorial equations.
Section \ref{s:results-1} describes our $GW$ results for benzene, a typical small molecule. 
Section \ref{s:compression} describes our algorithm for the compression of the screened Coulomb interaction
that is needed to treat larger molecules.
Section \ref{s:maintain-n3}  explains how $O(N^{3})$ scaling can be maintained for 
large molecules by alternatively compressing and decompressing the Coulomb interaction.
Section \ref{s:algorithm-summary} presents a summary of the entire algorithm for performing $GW$ calculations.
In section \ref{s:results-2} we test our method on naphthalene and anthracene, and our conclusions are presented
in section \ref{s:conclusion}.

\section{Elementary aspects of Hedin's $GW$ approximation}
\label{s:elementary-aspects-$GW$}

The one-electron Green's function of a many-body system has proved to be 
a very useful concept in condensed matter theory. It allows to compute 
the total energy, the electronic density and other quantities arising from one-particle operators.
The one-electron Green's function $G(\bm{r},\bm{r}',t)$
has twice as many spatial arguments as the electronic density but it remains
a far less complex object than the many-body wave function. 
Furthermore, Hedin has found  an exact set of equations for a finite
set of  correlation functions of which the one-electron Green's function
is the simplest element. This set of equations has not been solved so far for
any system whatsoever.
However, as a zeroth order starting point to his coupled equations,
Hedin suggested the very successful $GW$ approximation for the 
self-energy
$\Sigma(\bm{r},\bm{r}',t)$. This approximation describes the change
of the non interacting electron propagator $G_0(\bm{r},\bm{r}',t)$ due
to interactions among the electrons. With the help of a self-energy, one can
find the interacting Green's function from Dyson's equation
\begin{equation}
G=G_0+G_0\Sigma G_0+G_0\Sigma G_0\Sigma G_{0}+\ldots=\frac{1}{G_0^{-1}-\Sigma}, \label{Dyson}
\end{equation}
where products and inversions in the equation must be understood in an operator sense as 
required in many-body perturbation theory.

In Hedin's $GW$ approximation, the interaction of electrons with themselves
is taken into account by the following self-energy

\begin{equation}
\Sigma (\bm{r},\bm{r}',t)=\mathrm{i}G_{0}(\bm{r},\bm{r}',t)W(\bm{r},\bm{r}',t),
\label{self-energy}
\end{equation}%
where $W(\bm{r},\bm{r}',t)$ is a screened Coulomb interaction. 

The key idea of Hedin's $GW$ approximation \cite{Hedin} is to incorporate 
the screening of the Coulomb interaction from the very beginning in 
a zeroth order approximation.
Let $v(\bm{r},\bm{r}')=|\bm{r}-\bm{r}'|^{-1}$ be the bare Coulomb interaction and let
$\chi_0(\bm{r},\bm{r}',t-t')=\frac{\delta n(\bm{r},t)}{\delta V(\bm{r}',t')}$ 
be the density response $\delta n(\bm{r},t)$ of non interacting electrons with
respect to a change of the external potential $\delta V(\bm{r}',t')$.
Hedin then replaces the original Coulomb interaction  $v(\bm{r},\bm{r}')$ 
by the screened Coulomb interaction $W(\bm{r},\bm{r}', \omega)$ within
the random phase approximation (RPA)~\cite{RPA} 
\begin{equation}
W(\bm{r},\bm{r}',\omega)=\frac{1}{\delta(\bm{r}-\bm{r}''')-
v(\bm{r},\bm{r}'')\chi_0(\bm{r}'',\bm{r}''',\omega)}v(\bm{r}''',\bm{r}'),
\label{Coulombscreening}
\end{equation}
here and in the following we assume integration over
repeated spatial coordinates  (in our case $\bm{r}''$ and $\bm{r}'''$) on the right hand 
side of an equation
if they do not appear on its left hand side. This convention makes our equations more 
transparent without introducing ambiguities and it is analogous to the familiar Einstein's 
convention of summing over repeated indices. 

We can justify the expression (\ref{Coulombscreening}) by considering an internal screening
field $\delta V_{\mathrm{induced}}(\bm{r},\omega)$ that is generated by an extra external field
$\delta V_{\mathrm{external}}(\bm{r},\omega)$ 
\begin{equation}
\delta V_{\mathrm{total}}(\bm{r},\omega) =\delta V_{\mathrm{external}}(\bm{r},\omega)+\delta V_{\mathrm{induced}}(\bm{r},\omega),
\label{screening}
\end{equation}%
where
\begin{equation}
\delta V_{\mathrm{induced}}(\bm{r},\omega) =v(\bm{r},\bm{r}'')\delta n_{\mathrm{induced}}(\bm{r}'',\omega)
=v(\bm{r},\bm{r}'')\chi_0(\bm{r}'',\bm{r}''',\omega)\delta V_{\mathrm{total}}(\bm{r}''',\omega).
\nonumber
\end{equation}%
As a consequence we obtain a frequency dependent change of the total potential
\begin{equation}
\delta V_{\mathrm{total}}(\bm{r},\omega)=\frac{1}{\delta(\bm{r}-\bm{r}''')-
v(\bm{r},\bm{r}'')\chi_0(\bm{r}'',\bm{r'''},\omega)}\delta V_{\mathrm{external}}(\bm{r}''',\omega).
\label{RPA}
\end{equation}

If we assume that large fields are screened the same way as small field changes, then
we may replace $\delta V_{\mathrm{external}}(\bm{r},\omega)$
by the singular Coulomb interaction $v(\bm{r},\bm{r}')$ 
and we obtain the screened counterpart $W(\bm{r},\bm{r}',\omega)$ of the original 
bare Coulomb interaction as in eq (\ref{Coulombscreening}).

Because of the relation 
\begin{equation}
\mathrm{i}\chi_0(\bm{r},\bm{r}',t)=2G_{0}(\bm{r},\bm{r}',t)G_{0}(\bm{r}',\bm{r},-t)
\label{more_screening}
\end{equation}
the screening in eq (\ref{Coulombscreening}) may be interpreted as being due to
the creation of virtual electron-hole pairs. 
The screening by virtual electron-hole
pairs is the quantum analogue of classical Debye 
screening in polarizable media~\cite{Debyemodel}.
The factor of $2$ in eq~(\ref{more_screening}) takes
into account the summation over spins.

Many body theory uses Feynman-Dyson perturbation theory \cite{ManyBodyText}
and the latter is formulated in terms of time ordered correlators. For instance,
a Green's function is represented as a time ordered correlator 
of electron creation $\psi^{+}(\bm{r},t)$ and annihilation $\psi(\bm{r},t)$ operators 
\begin{equation}
\mathrm{i}G(\bm{r},\bm{r}',t-t') =
\theta (t-t')\langle 0|\psi(\bm{r},t)\psi^{+}(\bm{r}',t')|0\rangle-
\theta (t'-t)\langle 0|\psi^{+}(\bm{r}',t')\psi (\bm{r},t)|0\rangle,
\label{gf-via-creation-annihilation}
\end{equation}%
where the minus sign is due to Fermi statistics, $|0\rangle$ denotes
the electronic ground state, and where 
$\theta(t)$ denotes Heaviside's step function. This completes
our formal description of Hedin's $GW$ approximation. 

In practice, Hedin's equations are solved ``on top'' of 
a density functional or Hartree-Fock calculation. 
The framework of density functional theory (DFT)~\cite{HK,KS}
already includes electron correlations at the mean-field
level via the exchange correlation energy
$E_{\mathrm{\mathrm{xc}}}[n(\bm{r})]$, where $[n]$ denotes the functional
dependence of $E_{\mathrm{\mathrm{xc}}}$ on the electron density.
DFT calculations are usually performed using the
Kohn-Sham scheme~\cite{KS}, in which electrons move
as independent particles 
in an effective potential. The Kohn-Sham Hamiltonian $H_{\mathrm{KS}}$ reads
\begin{align}
H_{\mathrm{\mathrm{KS}}} &=-\frac{1}{2}\nabla^2+V_{\mathrm{KS}}, \\
V_{\mathrm{KS}} &=V_{\mathrm{ext}}+V_{\mathrm{Hartree}}+V_{\mathrm{xc}}
\text{, where }
V_{\mathrm{xc}}(\bm{r})=\frac{\delta E_{\mathrm{xc}}}{\delta n(\bm{r})}.  \notag
\end{align}%
To avoid including the interaction twice, the exchange correlation potential 
$V_{\mathrm{xc}}(\bm{r})$
must be subtracted from $\Sigma(\bm{r},\bm{r}',t)$ in eq (\ref{self-energy})
when using the output of a DFT calculation as input for a $GW$ calculation.
This is done by making the replacement
\begin{equation}
\Sigma(\bm{r},\bm{r}',t) \rightarrow \Sigma(\bm{r},\bm{r}',t) -
\delta(\bm{r}-\bm{r}')\delta(t) V_{\mathrm{xc}}(\bm{r})
\end{equation}%
in Dyson's equation (\ref{Dyson}).

Our aim is to compute the electronic density of states (DOS)
that is defined as the trace of the 
imaginary part of the electron propagator 
\begin{equation}
\rho(\omega +\mathrm{i}\varepsilon )=-\frac{1}{\pi}
\mathrm{Im}\int G(\omega +\mathrm{i}\varepsilon,\bm{r},\bm{r})d^3r.
\end{equation}%
The electronic DOS 
$\rho (\omega +\mathrm{i}\varepsilon )$ can be compared with experimental
data from direct and inverse photo-emission~\cite{PhotoEmissionExperiments}. 
From it, we can read off the energy position of the highest occupied and 
the lowest unoccupied molecular orbitals (HOMO and LUMO)  
or, alternatively, the ionization energy and the electron affinity.

Finally, let us list the equations that define the $GW$ 
approximation
\begin{equation}
\begin{aligned}
\mathrm{i}\chi_0(\bm{r},\bm{r}',t) &=
2 G_0(\bm{r},\bm{r}',t)G_0(\bm{r}',\bm{r},-t);       & \text{free electron response}  \\
W(\bm{r},\bm{r}',\omega ) &=\left[\delta(\bm{r}-\bm{r}''')
-v(\bm{r},\bm{r}'')\chi_0(\bm{r}'',\bm{r}''',\omega )\right]^{-1}
v(\bm{r}''',\bm{r}');                              & \text{RPA screening}  \\
\Sigma (\bm{r},\bm{r}',t) &= 
\mathrm{i}G_0(\bm{r},\bm{r}',t)W(\bm{r},\bm{r}',t); &\text{$GW$ self-energy}  \\
G^{-1}(\bm{r},\bm{r}',\omega +\mathrm{i}\varepsilon ) &=G_0^{-1}(\bm{r},\bm{r}',
\omega +\mathrm{i}\varepsilon )-\Sigma (\bm{r},\bm{r}',\omega +\mathrm{i}\varepsilon ).
&\text{Dyson equation}
\end{aligned}
\label{all_GW_equations}
\end{equation}%

The next two sections describe 
the tensor form of equations (\ref{all_GW_equations}) as well as 
the main ingredients of our
implementation of the $GW$ approximation as embodied 
in equations (\ref{all_GW_equations}) for
the case of small molecules. Later sections will describe the compression/decompression
of the Coulomb interaction that is needed for treating 
large molecules without over flooding the computer memory.

\section{Tensor form of Hedin's equations}
\label{s:tensor-form}

In order to compute the non interacting Green's function (\ref{gf-via-creation-annihilation}),
we will use the LCAO method 
where one expresses the electron operator in terms of a set of
fermions $c_{a}(t)$ that belong to localized atomic orbitals \cite{Fulde} 
\begin{equation}
\psi(\bm{r},t)\sim \sum_{a}f^{a}(\bm{r})c_{a}(t).  \label{Fulde}
\end{equation}%
Such a parametrization is parsimonious in the number of degrees of freedom,
although its quality is difficult to control and to improve in a systematic way.

The output of a DFT calculation (that serves as input for the $GW$ calculation)
is the Kohn-Sham Hamiltonian $H^{ab}$ and the overlap matrix $S^{ab}$ of 
the LCAO basis functions $f^{a}(\bm{r})$ \cite{Martin}.
One may use the eigenvectors $\{X_{a}^{E}\}$ of the Kohn-Sham Hamiltonian 
\begin{equation}
H^{ab}X_{b}^{E}=ES^{ab}X_{b}^{E} \label{KS_equation}
\end{equation}%
to express the (time-ordered) propagation of electrons between localized atomic orbitals

\begin{equation}
G_{ab}^0(\omega \pm \mathrm{i} \varepsilon)= \sum_E \frac{X_a^E X_b^E}
{\omega \pm \mathrm{i} \varepsilon -E}. \label{local_propagator}
\end{equation}%\mathrm{i}

In this paper we measure energies relative to a Fermi energy, so that
$E<0$ $(E>0)$ refers to occupied (empty) states, respectively,
and the infinitesimal constant $\varepsilon$ shifts the poles of the Greens function
away from the real axis. Moreover, to avoid cluttering up the notation, we will often use
Einstein's convention of summing over repeated indices, as in eq (\ref{KS_equation}).

The set of equations (\ref{all_GW_equations}) contains correlation functions,
such as the
density response function $\chi(\bm{r}, \bm{r}', t)$
that must be represented in a basis of \textit{products of atomic orbitals}
\cite{ManyBodyText}

\begin{equation}
\mathrm{i}\chi(\bm{r},\bm{r}',t-t') =\theta(t-t')
\langle 0|n(\bm{r},t)n(\bm{r}',t')|0\rangle
+\theta(t'-t)\langle 0|n(\bm{r}',t')n(\bm{r},t)|0\rangle.
\end{equation}
Indeed, by virtue of eq (\ref{Fulde}), the electronic density
$n(\bm{r},t)=\psi^{+}(\bm{r},t)\psi(\bm{r},t)$ involves products
of atomic orbitals
\begin{equation*}
n(\bm{r},t)=\psi^{+}(\bm{r},t)\psi(\bm{r},t)=
\sum_{a,b}f^a(\bm{r})f^b(\bm{r})c_a^{+}(t)c_b(t).
\end{equation*}

The  set of products $\{f^a(\bm{r})f^b(\bm{r})\}$ is well known to be strongly
linearly dependent \cite{LinearDependenceOldpaper}. As an improved solution
of this very old technical difficulty, we previously 
developed an algorithm to construct a local basis of ``dominant products''
$F^{\mu}(\bm{r})$ that
\textit{i}) spans the space of orbital products with exponential accuracy and which
\textit{ii}) respects the locality of the original atomic orbitals \cite{DF}. Moreover, the products 
of atomic orbitals $f^a(\bm{r})f^b(\bm{r})$ relate to dominant products
$F^{\mu}(\bm{r})$ via a \textit{product vertex} $V_{\mu }^{ab}$ 
\begin{equation}
f^a(\bm{r})f^b(\bm{r})=\sum_{\mu }V_{\mu }^{ab}F^{\mu }(\bm{r}).
\label{VertexDefinition}
\end{equation}%

Because the dominant products $F^{\mu}(\bm{r})$ are themselves special linear
combinations of the original products, arbitrary extra fitting functions do
not enter into this scheme. In order to respect the principle of locality,
the above decomposition is carried out separately for each pair of atoms,
the orbitals of which overlap. 
By their construction, the set of coefficients $V_{\mu }^{ab}$
is sparse in the sense that $V _{\mu }^{ab}\neq
0$ only if $a,b,\mu $ all reside on the same atom pair \cite{DF}. 
In the construction of the dominant product basis, 
we made use of Talman's algorithms and computer codes for the expansion of products of
orbitals about an arbitrary center and we also used his fast Bessel transform~\cite{Talman}.

To rewrite the defining equations of the $GW$ approximation (\ref{all_GW_equations})
in our basis, we expand both $G(\bm{r},\bm{r}',t-t')$
and $\Sigma (\bm{r},\bm{r}',t-t')$ in atomic orbitals
$f^a(\bm{r})$ \cite{PK+DF+OC} 
\begin{equation}
\begin{aligned}
G(\bm{r},\bm{r}',t-t') &=  G_{ab}(t-t')f^a(\bm{r})f^b(\bm{r}'); \\
\Sigma (\bm{r},\bm{r}',t-t') &= \Sigma_{ab}(t-t')f^a(\bm{r})f^b(\bm{r}').
\end{aligned}
\label{expansion1}
\end{equation}
We also develop the screened Coulomb interaction in dominant products
\begin{equation}
W^{\mu \nu }(t-t')=\int d^3r d^3r'F^{\mu }(\bm{r})W(\bm{r},\bm{r'},t-t')F^{\nu }(\bm{r}').
\label{expansion2}
\end{equation}%
Using eqs (\ref{VertexDefinition}, \ref{expansion1}, \ref{expansion2}) 
it is easy to show \cite{PK+DF+OC}
that Hedin's equations (\ref{all_GW_equations}) take the following tensorial form in our basis
\begin{align}
\mathrm{i}\chi_{\mu\nu}^0(t) &= 2 V _{\mu }^{aa'}G_{ab}^0(t)V _{\nu }^{bb'}G_{a'b'}^0(-t);
&\text{free electron response} \label{tensorform-response}  \\
W^{\mu \nu }(\omega ) &=\frac{1}{\delta_{\alpha }^{\mu }-v^{\mu \beta }
\chi_{\beta\alpha}^{0}(\omega)}v^{\alpha\nu};
&\text{RPA screening}  \label{tensorform-scr-inter} \\
\Sigma^{ab}(t) &=\mathrm{i}V_{\mu}^{aa'}G_{a'b'}^0(t)V_{\nu}^{b'b} W^{\mu \nu }(t);
&\text{$GW$ approximation}  \label{tensorform-self-energy} \\
G_{ab}^{-1}(\omega +\mathrm{i}\varepsilon ) &=
G_{0ab}^{-1}(\omega +\mathrm{i}\varepsilon )-\Sigma_{ab}(\omega +\mathrm{i}\varepsilon).
&\text{Dyson's equation}
\label{tensorform-dyson}
\end{align}%
Here $v^{\mu \nu }$ denotes the Coulomb interaction 
$v^{\mu \nu }=\int d^3r d^3r'\, F^{\mu }(\bm{r})\frac{1}{|\bm{r}-\bm{r}'|}F^{\nu }(\bm{r}')$
which, due to its positivity and symmetry, we also refer to as a ``Coulomb metric''.
Indices are raised or lowered using either the overlaps of the dominant
functions $F^{\mu }(\bm{r})$ or the overlaps of the 
atomic orbitals $f^a(\bm{r})$ 
and which are defined as follows
\begin{equation}
O^{\mu \nu }=\int d^3r\, F^{\mu }(\bm{r})F^{\nu }(\bm{r}), \ 
S^{ab}=\int d^3r\, f^{a}(\bm{r})f^{b}(\bm{r}).
\end{equation}%

\begin{figure}
\centerline{\includegraphics[width=7cm, angle=0,clip]{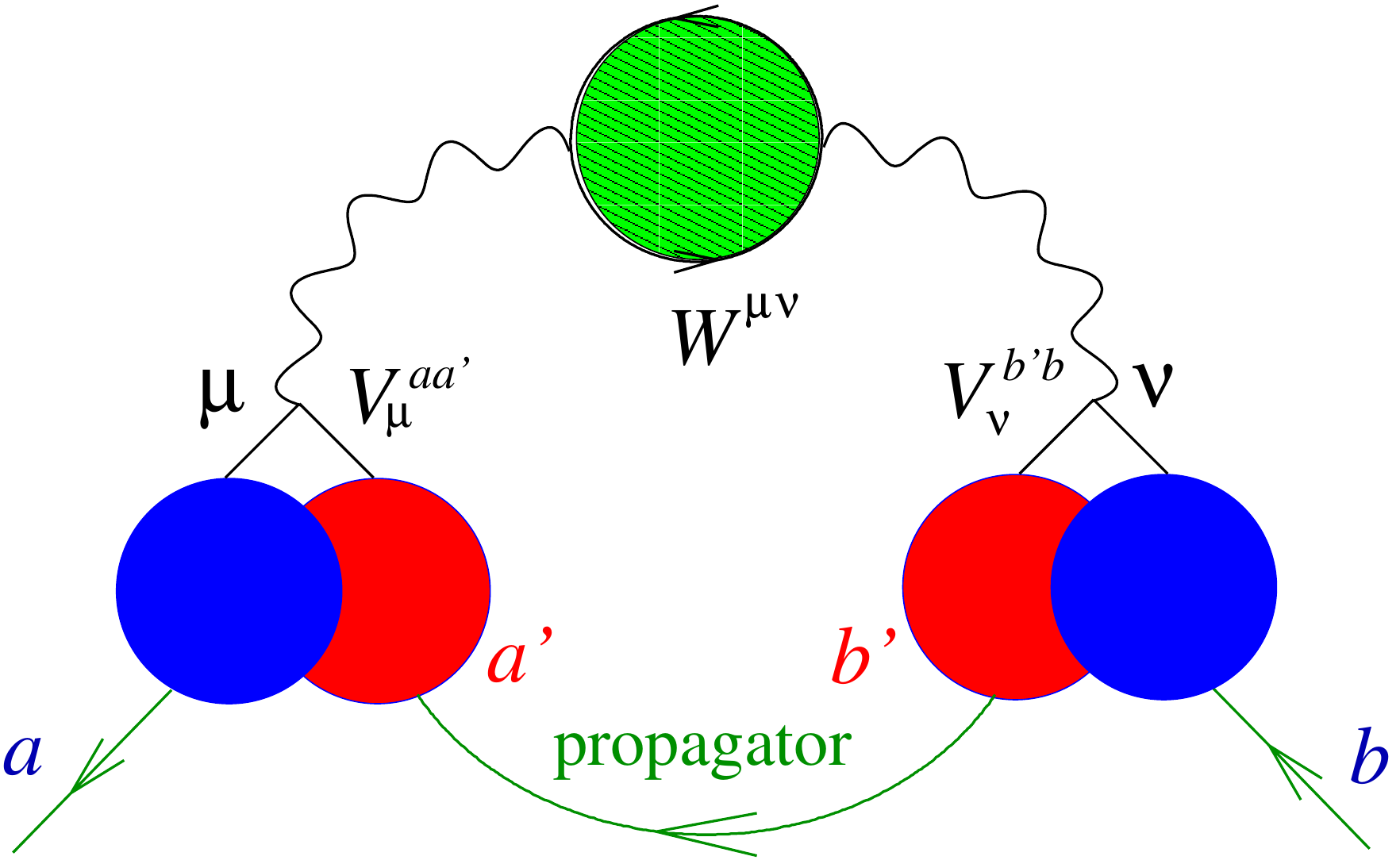}}
\caption{Feynman diagram for the $GW$ self-energy
expressed in our local LCAO and dominant products basis. 
\label{f:self-energy-diagram}}
\end{figure}

Figure~\ref{f:self-energy-diagram} shows the Feynman diagram corresponding
to eq~(\ref{tensorform-self-energy}). The local character of the product
vertex $V_{\mu}^{aa'}$ is emphasized in this figure.

\section{The instantaneous part of the self-energy }
\label{s:instant-scr-inter}

When the screened Coulomb interaction $W^{\mu \nu }$ in 
eq (\ref{tensorform-scr-inter}) is expanded as a function of
$v \chi^0$, its first term is the bare
Coulomb interaction $v^{\mu \nu}\delta (t-t')$ 
and the corresponding self-energy in eq~(\ref{tensorform-self-energy}) is
frequency independent. In textbook treatments of the 
theory of the electron gas, it is explained \cite{ManyBodyText}
that the Green's function $G_{ab}(t-t')$ of the electron gas 
must be defined, at $t=t'$, by setting $t-t'=0^{-}$ or by first annihilating
and then creating electrons. Using this prescription and eqs
(\ref{tensorform-scr-inter}, \ref{tensorform-self-energy}) one finds
the following result for the frequency-independent self-energy
that corresponds to the exchange operator

\begin{equation}
\Sigma_{\mathrm{x}}^{ab}(t-t')=
\mathrm{i}V _{\mu }^{aa'}G_{a'b'}^0(0^{-})V_{\nu }^{b'b} \delta(t-t') v^{\mu\nu} \nonumber.
\end{equation}
In the frequency domain, the last operator becomes a frequency independent matrix
\begin{equation}
\Sigma_{\mathrm{x}}^{ab}= V_{\mu}^{aa'} \sum_{E<0} X^{E}_{a'} X^{E}_{b'} V_{\nu}^{b'b} v^{\mu\nu},
\label{instantaneous}
\end{equation}%
which can be computed in $O(N^3)$ operations by using the sparsity
of the product vertex  $V_{\mu}^{aa'}$.

For small molecules and clusters, the instantaneous self-energy
that incorporates the effects of electron exchange may dominate over
the remaining frequency dependent self-energy. If this is the case,
we may substitute $\Sigma_{\mathrm{x}}^{ab}$ into
eq (\ref{tensorform-self-energy}) and finish the 
calculation by computing the DOS 
 
$$
\rho(\omega +\mathrm{i}\varepsilon ) =
-\frac{1}{\pi } S^{ab}\,\mathrm{Im}G_{ba}(\omega +\mathrm{i}\varepsilon ),
$$
where we have emphasized the non-orthogonality of the basis orbitals
by the explicit inclusion of the overlap $S^{ab}$.

However, the frequency dependent part of the self-energy contains correlation 
effects that significantly improve the calculation 
quantitatively and  qualitatively as we demonstrate in sections
\ref{s:results-1} and \ref{s:results-2}. 
Therefore, we shall present our approach for the frequency dependent part 
of the self-energy in the next section. 

\section{Using spectral functions to compute the self-energy}
\label{s:sf-techniques}

One might want to solve the eqs (\ref{tensorform-scr-inter}) and (\ref{tensorform-dyson})
directly as matrix valued equations in time $t$ and to use FFTs
\cite{NumericalRecipesFFT} to shuttle back and forth between the time and
frequency domains. Unfortunately, however, this direct approach is
doomed to fail---the functions $\{G_{ab}(t), \Sigma ^{ab}(t), W^{\mu\nu}(t)\}$
are too singular at $t=0$ to be multiplied together.
We will now show how spectral functions come to the rescue and allow us to
\textit{i}) respect locality in our calculations and to 
\textit{ii}) accelerate our calculation 
by means of FFT.

Let us consider the energy dependent density matrix
\begin{equation}
\rho _{ab}(\omega )=\sum_{E}X_{a}^{E}X_{b}^{E}\delta (\omega -E)
\end{equation}%
and rewrite the electronic propagator (\ref{local_propagator}) with the help of it.
\begin{equation}
G_{ab}^{0}(\omega \pm i\varepsilon )=
\int_{-\infty }^{\infty }ds\,\frac{\rho _{ab}(s)}{\omega \pm i\varepsilon - s}.
\label{gf-via-sf-freq}
\end{equation}%
Integral representations such as these are very useful, even in finite
systems where the spectral weight is concentrated at isolated 
frequencies~\cite{DF+PK,PK+DF+OC}.
Because a spectral function is broadened by the
experimental resolution $\varepsilon$, it can be 
represented on a discrete mesh of
frequencies, with the distance between mesh points 
somewhat smaller than $\varepsilon$.
All the response functions considered in the present paper
have a spectral representation because their retarded and advanced parts
taken together define a single analytic function in the complex frequency
plane with a cut on the real axis.

A spectral representation is merely a rather thinly disguised Cauchy
integral as we can see by considering the Cauchy integral representation of the
electronic Greens function%
\begin{equation}
G_{ab}(z)=\frac{1}{2\pi \mathrm{i}}\oint_{C}\frac{G_{ab}(\xi)d\xi }{\xi -z},
\end{equation}%
where $C$ is a path surrounding the point $z$ with $\mathrm{Im}z>0$ 
in an anti-clockwise direction. 
If the point $z=\infty $ is regular, the complex 
plane may be treated like a sphere and we may deform 
the contour on this sphere in such a way that it wraps
around the cut on the real axis in a clockwise direction. 
Finally, because Green's functions take mutually hermitian conjugate values 
$G_{ab}(z^{\ast })=G^{\ast }_{ba}(z)$ across
the cut on the real axis, the above integral can the rewritten as 
\begin{equation}
G_{ab}(z)=\int ds\frac{\rho _{ab}(s)}{z-s}\text{, }
\rho _{ab}(z)=-\frac{1}{\pi }\mathrm{Im} G_{ab}^{0}(z)
\end{equation}%
with $z$ on the upper branch of the cut. 
In writing the preceding equation we have used the simplifying feature that
the electronic Green's function is
a symmetric matrix in our real representation of angular momenta (the same is
true for the screened Coulomb interaction).
In the following, we will always reconstruct correlation functions such as
$\{G_{ab}(\omega ),\,\Sigma ^{ab}(\omega ),\,W^{\mu \nu }(\omega )\}$ from their
imaginary part or from their spectral functions. The time ordered version of such
correlators  is determined above (below) the real axis for positive
(negative) frequencies, respectively.

\subsection{The spectral function of a product of two correlators}

The well known convolution theorem \cite{NumericalRecipesFFT} tells us
that the spectral content of a product of two signals is the convolution
of the spectral contents of its factors. The situation is entirely analogous
for Green's functions and the other correlators considered here
and their spectral functions. 
To see this, we use the spectral representations of the time ordered factors
$G_{ab}(t)$, 
$\Sigma ^{ab}(t)$,
$W^{\mu \nu }(t)$
(the quantities in eqs (\ref{tensorform-response}--\ref{tensorform-dyson})
are time ordered) 
\begin{equation}
\begin{aligned}
G_{ab}(t) &=
-\mathrm{i}\theta(t)\int_{0}^{\infty }ds\,\rho_{ab}^{+}(s)e^{-\mathrm{i}st}
+\mathrm{i}\theta(-t)\int_{-\infty }^{0}ds\,\rho_{ab}^{-}(s)e^{-\mathrm{i}st}; \\
\Sigma ^{ab}(t) &=
-\mathrm{i}\theta(t)\int_{0}^{\infty }ds\,\sigma_{+}^{ab}(s)e^{-\mathrm{i}st}
+\mathrm{i}\theta(-t)\int_{-\infty }^{0}ds\sigma_{-}^{ab}(s)e^{-\mathrm{i}st}; \\
W^{\mu \nu }(t) &=
-\mathrm{i}\theta(t)\int_{0}^{\infty }ds\,\gamma_{+}^{\mu \nu}(s)e^{-\mathrm{i}st}
+\mathrm{i}\theta(-t)\int_{-\infty }^{0}ds\gamma_{-}^{\mu \nu}(s)e^{-\mathrm{i}st},
\end{aligned}
\label{spectral_1}
\end{equation}
where ``positive'' and ``negative'' spectral functions 
define the whole spectral function by means of Heaviside functions. For instance,
the spectral function of the electronic Green's function reads
$$
\rho_{ab}(s)=\theta(s)\rho^{+}_{ab}(s)+\theta (-s)\rho^{-}_{ab}(s).
$$

These representations can be checked by transforming (for example)
the representation of $G_{ab}(t)$ into the frequency domain 
and by comparing with the known expression (\ref{gf-via-sf-freq}).
We then compute $\Sigma^{ab}(t)$ from eq (\ref{tensorform-self-energy}) and
compare the result with the second of eqs (\ref{spectral_1})
The spectral function of the self-energy is seen to have the expected convolution form 
\begin{align}
\sigma_{+}^{ab}(s) &=\int_{0}^{\infty }\,\int_{0}^{\infty }
\delta (s_{1}+s_{2}-s)\,V_{\mu }^{aa'}\rho _{a'b'}^{+}(s_{1})
V_{\nu }^{b'b}\gamma_{+}^{\mu \nu }(s_{2})ds_1 ds_2,  \label{spectral_3} \\
\sigma_{-}^{ab}(s) &=-\int_{-\infty }^{0}\,\int_{-\infty}^{0}
\delta (s_{1}+s_{2}-s)V _{\mu }^{aa'}\rho _{a'b'}^{-}(s_{1})
V_{\nu }^{b'b}\gamma _{-}^{\mu \nu}(s_{2})ds_1 ds_2.  \notag
\end{align}%
Note that, as commented above, the $V_{\mu }^{aa'}$ matrices are sparse 
and respect spatial locality.
Finally, we can easily construct the full self-energy from its spectral functions
$\sigma_{\pm }^{ab}(s)$ by a Cauchy type integral 
\begin{equation}
\Sigma ^{ab}(\omega \pm i\varepsilon )=\int_{-\infty }^{\infty }\frac{\sigma ^{ab}(s)ds}{%
\omega \pm i\varepsilon -s}.  \label{Cauchy1}
\end{equation}%

By entirely analogous arguments, we can find the spectral function of the non
interacting response $\chi_{\mu \nu }^{0}$ from eq (\ref{tensorform-response})
\begin{equation}
\begin{aligned}
a_{\mu \nu }(s) &=\int_{0}^{\infty } \int_{0}^{\infty }%
V _{\mu }^{ab}\rho _{bc}^{+}(s_{1})V _{\nu }^{cd}
\rho_{da}^{-}(-s_{2})\delta (s_{1}+s_{2}-s)ds_{1}ds_{2}, &\text{for }s>0;
\\
a_{\mu \nu }(-s) &= -a_{\mu \nu }(s), &\text{for all } s; \\
\chi _{\mu \nu }^{0}(\omega +\mathrm{i}\varepsilon ) &=\int_{-\infty }^{\infty }ds\,%
\frac{a_{\mu \nu }(s)}{\omega +\mathrm{i}\varepsilon -s}, & \text{for }\omega >0. 
\end{aligned}%
\label{chi_by_spectra}
\end{equation}
We implemented the convolutions in eqs (\ref{spectral_3}, \ref{chi_by_spectra})
conveniently by FFT without encountering any singularities.
Please observe that analytic continuations are not needed in our approach.

We have seen in this subsection that the locality of the expressions for
$\Sigma ^{ab}(t)$ and $\chi_{\mu \nu }^{0}(t)$ in eqs
(\ref{tensorform-response}) and (\ref{tensorform-self-energy})
can be taken into account without multiplying singular Green's functions and 
by focusing instead on the spectral functions of their products.

\subsection{ The second window technique}

\label{s:two-window-technique}
Although we only need results in a suitable low energy window
$-\lambda \leq \omega \leq \lambda $ of a few electron volts, 
eqs (\ref{Cauchy1}, \ref{chi_by_spectra}) show that high energy processes
at $|\omega |>|\lambda |$ influence 
quantities at low energies, such as,
for example, the self-energy. Therefore, these
high energy processes cannot be ignored and we need the imaginary part
of the screened Coulomb interaction $W^{\mu \nu }$
not only for small $|\omega |\leq \lambda $ but also for larger frequencies.
To find the imaginary part of $W^{\mu \nu }$, we also need, in view of
eq (\ref{tensorform-self-energy}), the non interacting response
$\chi_{\mu\nu}^{0}$ both at small and at large frequencies.

Let us see in the case of the density response, how the necessary spectral
information can be obtained from two separate calculations in two distinct
frequency windows \cite{DF+PK}. In the large spectral window
$-\Lambda \leq \omega \leq \Lambda $ a low resolution calculation with a
large broadening 
(and, therefore, a coarse grid of frequencies)
is sufficient to find $\chi _{\mu \nu }^{0}$ at large
energies $|\omega |>|\lambda |$ 
\begin{equation}
\chi _{\mu \nu }^{0}(\omega +\mathrm{i}\varepsilon _{\text{large}})=\int_{-\Lambda
}^{\Lambda }ds\frac{a_{\mu \nu }(s)}{\omega +\mathrm{i}\varepsilon_{\text{large}}-s}.
\end{equation}%
To get correct results in the low energy window
$-\lambda \leq \omega \leq \lambda $ we must take into account the spectral
weight in this window
\begin{equation}
\begin{aligned}
\chi _{\mu \nu }^0(\omega +\mathrm{i}\varepsilon _{\text{small}}) &=\int_{-\lambda
}^{\lambda }ds\frac{a_{\mu \nu }(s)}{\omega +\mathrm{i}\varepsilon _{\text{small}}-s}%
\ +\left( \int_{-\Lambda }^{-\lambda }+\int_{\lambda }^{\Lambda }\right) ds%
\frac{a_{\mu \nu }(s)}{\omega +\mathrm{i}\varepsilon _{\text{large}}-s} \\
&=\chi _{\mu \nu }^{\text{small window}}(\omega +\mathrm{i}\varepsilon _{\text{%
small}})+\left[ \chi _{\mu \nu }^{\text{large window}}(\omega +\mathrm{i}\varepsilon
_{\text{large}})\right] _{\text{truncated spectral function}}.
\end{aligned}
\label{win2-for-response}
\end{equation}
Instead of doing the second Cauchy integral directly, we construct
$\chi_{\mu \nu }^{\text{large window}}$ from the large spectral window, using
spectral data that are truncated for $|s|<\lambda $ to avoid counting
the spectral weight in $-\lambda \leq \omega \leq \lambda $ twice. Moreover, 
the broadening constant $\varepsilon$ is set differently in
the first and second windows and corresponds to the spectral resolution
in these windows. The broadening constant $\varepsilon$ is chosen
automatically in each frequency window, by setting $\varepsilon=1.5 \Delta \omega$,
where $\Delta \omega$ is the distance between two points on the 
corresponding frequency grid.
We use the second window technique (presented in eq (\ref{win2-for-response})
for the case of the density response) again in the calculation of the
self-energy $\Sigma^{ab}(\omega)$, where we also need the screened interaction
in two windows. We combine the spectral functions of the self-energy in
exactly the same way as for the density response. 
For the cases considered here, computations
using two spectral windows were up to one order of magnitude
faster than computations using a single spectral window.

\section{Testing our implementation of $GW$ on a small molecule \label{s:results-1}}

The methods presented above are sufficient to compute the self-energy
(\ref{self-energy}) of small molecules~\cite{comp-details}. 
As a test, we will compute the interacting
Green's function  by solving Dyson's equation (\ref{Dyson}). 
From this Green's function we can obtain the DOS and estimate
the positions of the HOMO and LUMO levels. 
Here we illustrate this procedure in the case of benzene.
This molecule has been chosen because extensive theoretical results
and experimental data are available for it.
Our calculations show a considerable improvement 
using the $GW$ approximation 
as compared to the results obtained with plain DFT calculations using local or
semi-local functionals.
In general, for small molecules we find a reasonable agreement with
experimental data and previous $GW$ calculations 
of the ionization potentials and electron affinities.

The input for our $GW$ method has 
been obtained from calculations using the local density approximation (LDA) 
and the SIESTA package~\cite{siesta}. 
SIESTA uses a basis set of strictly confined numerical atomic orbitals.
The extension of these orbitals is
consistently determined by
an \textit{energy shift} parameter. In general, the smaller the energy shift 
the larger the extension of the orbitals, although the procedure
results in different cutoff radii 
for each multiplet of orbitals~\cite{Artacho-about-Energy-Shift}.
In the present calculations we have used the default 
double-$\zeta$ polarized (DZP) basis, along with 
the Perdew-Zunger LDA exchange-correlation functional~\cite{Perdew-Zunger}
and pseudo-potentials
of the Troullier-Martins type~\cite{Troullier-Martins}. 
Our calculations indicate (see table~\ref{t:ip-ea-benzene}) that it is necessary to 
use rather extended orbitals to obtain converged 
results for the HOMO and LUMO levels.
For the most extended basis used here 
(determined from an energy shift of 3~meV) all the 
orbitals in benzene have a non-zero overlap and, in principle, the number
of products of orbitals is 108(108+1)/2=5886. This number is 
reduced  
using the algorithm described in Ref.~\cite{DF}, and the  
dominant product basis (see eq~(\ref{VertexDefinition}) )
only contains 2325 functions. 
The spectral functions have been discretized 
using a grid with $N_{\omega}=1024$ points
in the range from $-80$ eV to $80$ eV. The broadening constant has been
set automatically to $\varepsilon=1.5\Delta \omega=0.234375$ eV.
The frequency range was chosen manually by inspecting the 
non interacting absorption spectrum. The results of the calculation 
depend only weakly on the frequency range.

\begin{figure}[htb]
\centerline{
\begin{tabular}{m{7cm}m{0.1cm}m{5cm}}
\centerline{
\includegraphics[width=7cm, viewport=50 50 410 300, angle=0,clip]{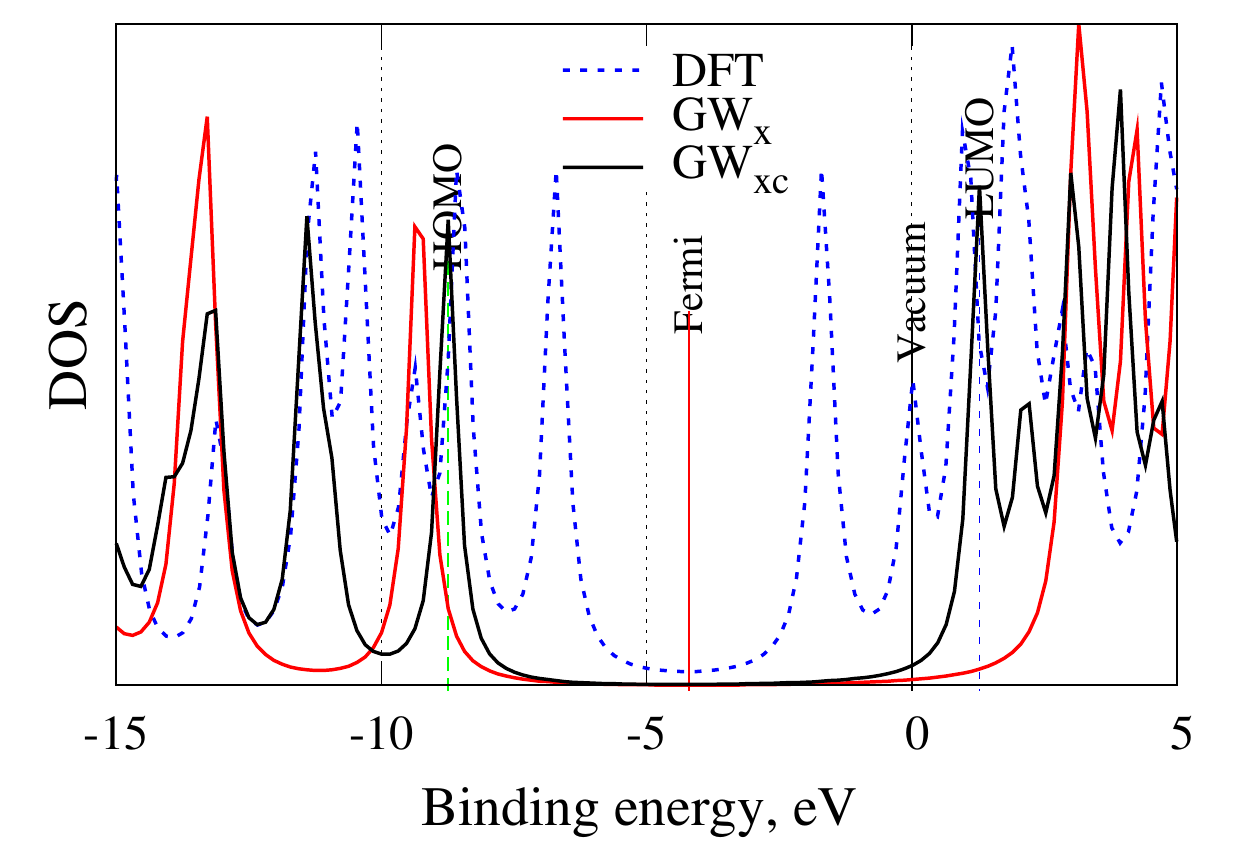}} & &
\centerline{\includegraphics[width=4.5cm,angle=0,clip]{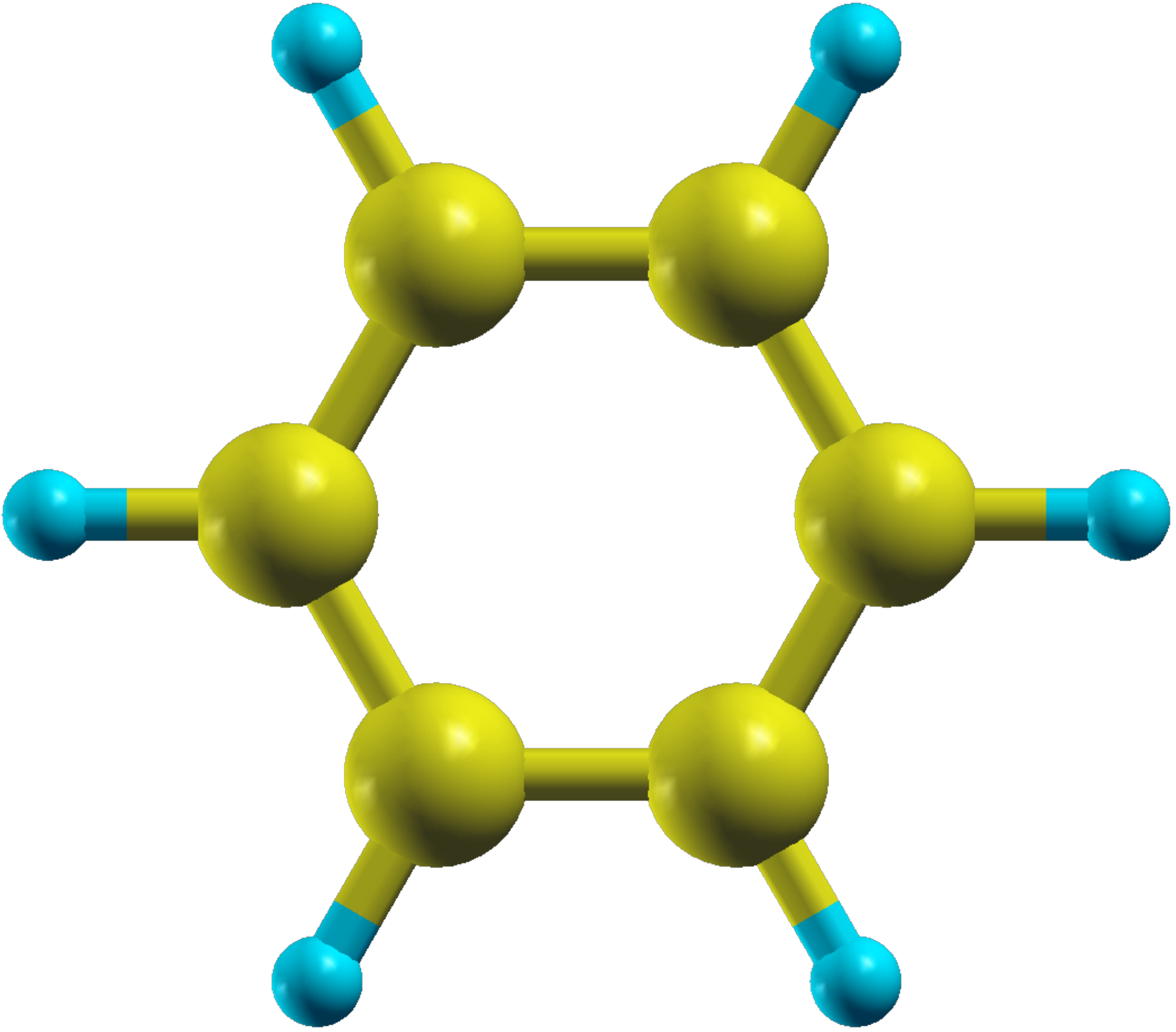}} \\[-0.2cm]
\centerline{a)} & & \centerline{b)}
\end{tabular}}
\caption{a) Density of states of benzene computed from different
Green's functions using as an input the 
results of a DFT-LDA calculation performed with the SIESTA
package. A DZP basis set, with orbital radii determined
using  a value of the  energy shift parameter of 3~meV, has been used.
The results shown in this figure are obtained with a single energy
window. $GW_{\mathrm{x}}$ refers to the results obtained with only
the instantaneous part of the self-energy (only exchange), 
while $GW_{\mathrm{xc}}$ labels the results obtained with the
whole self-energy (incorporating additional correlation effects).
b) Ball and stick model of benzene produced with the 
XCrysDen package \cite{xcrysden}.
\label{f:benzene-dos}}
\end{figure}

Figure \ref{f:benzene-dos} shows the DOS
calculated with different Green's functions. 
As one can see, the input Green's function $G_0$ 
from a DFT-LDA calculation has
a very small HOMO-LUMO gap. The Green's function $G$ obtained 
with the instantaneous part of
the self-energy (see eq~\ref{instantaneous}) opens the HOMO-LUMO gap.
This part of the self-energy $\Sigma_{\mathrm{x}}$ incorporates the effect
of exchange and is very important for small molecules. However,
the gap is over-estimated as one can already anticipate from typical
mean-field Hartree-Fock calculations.
Correlation effects are partially taken into account by
the  dynamical part of the $GW$ self-energy. This brings the
HOMO-LUMO gap closer to the experimental value. 
Our results stay also in agreement with other works using
similar approximations ($G_0W_0$ on top of DFT-LDA)
\cite{Baroni$GW$,Tiago-Chelikowsky-and-Frauenheim}.

Apart from the $GW$ approximations to the self-energy (\ref{self-energy}),
our numerical method is controlled by precision parameters 
of a more technical nature. 
Table~\ref{t:ip-ea-benzene} present the results for the ionization 
(IP) and electron affinity 
(EA) as a function of the 
extension of the atomic orbitals in the original LDA calculation
as determined from the energy shift 
parameter~\cite{Artacho-about-Energy-Shift}.
An energy shift of 150~meV is usually sufficient to have
an appropriate description of the ground-state properties
of the molecules~\cite{siesta}. However, we can see 
that our $GW$ calculation requires more extended (smaller energy shift)
orbitals. 
The slow convergence of the ionization potential of benzene
with respect to the quality/completeness of the basis set
has also been observed in the plane-wave calculations (using
Wannier functions) of Ref.~\cite{Baroni$GW$}.

\begin{figure}[htb]
\centerline{\includegraphics[width=7cm,viewport=50 50 410 300, angle=0,clip]{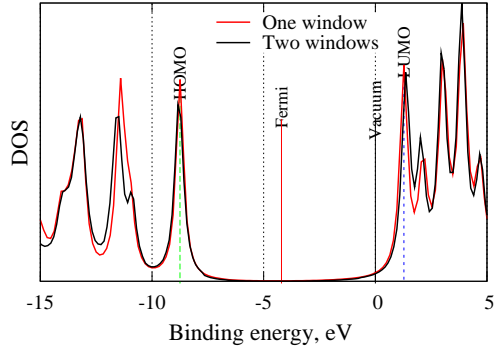}}
\caption{The density of states of benzene computed with a uniformly discretized
spectral function and using the second window technique. The peak positions are
very weakly perturbed by using the two windows technique. The parameters of the
calculation are identical with those of figure \ref{f:benzene-dos}.
The two windows technique allowed to reduce the number of frequency
points from $N_{\omega}=1024$ to $N_{\omega}=192$.
\label{f:benzene-dos-win1-win12}}
\end{figure}

Table \ref{t:ip-ea-benzene}
also shows the results of calculations using 
one and two energy windows.
The former calculation is more straightforward but requires
the same density of frequency points at higher energies as in the region
of interest near the HOMO and LUMO levels.
The latter calculation  uses two separate frequency grids:
as described in section \ref{s:two-window-technique}
a lower resolution and a larger imaginary part of the energy are
used for the whole spectral range, while
high resolution and a small width are used
in the low energy range to resolve the HOMO and LUMO levels. 
Thus, the second window technique 
requires the computation of both the response
function and screened Coulomb interaction at far fewer frequencies than
the one-window calculation.
For instance, the one-window results presented above have been
obtained with $N_{\omega}=1024$ frequencies, while the two-window
results used only $N_{\omega}=192$ frequencies in both windows,
implying a gain of a factor 2.7 in speed and in memory.
The first and second window extend to 12.58 eV and 80 eV, respectively.
The first window is chosen as
$2.5 (E_{\text{DFT LUMO}}-E_{\text{DFT HOMO}})$. 
The computational result depends only
weakly on the extension of the first window. The second window is chosen
manually as in the one-window calculation above. 
The broadening constant $\varepsilon$ has been set separately for
each spectral window, using the default value $\varepsilon=1.5\Delta \omega$.
The lower number of frequencies obviously accelerates the calculation
and saves memory, while introducing very small inaccuracies in the
low frequency region. According to table~\ref{t:ip-ea-benzene} the positions
of HOMO and LUMO agree within $0.1$~eV.
Figure~\ref{f:benzene-dos-win1-win12} shows that the second window 
leads to changes that are small, both in the HOMO and LUMO positions,
and in the density of states at low energies.

\begin{table}\centerline{
\begin{tabular}{|c|c|c||c|c|}
\hline
Energy-shift, & \multicolumn{2}{c||}{One window} & \multicolumn{2}{c|}{Two windows} \\
meV & IP, eV & EA, eV & IP, eV & EA, eV \\
\hline
150               &  8.48     & -1.89     &  8.48  & -2.01  \\
30                &  8.71     & -1.45     &  8.72  & -1.57  \\
3                 &  8.76     & -1.29     &  8.78  & -1.41  \\
\hline
Experiment        &  9.25     & -1.12     &  9.25  & -1.12  \\
\hline
\end{tabular}}
\caption{
Ionization potentials and  electron affinities for
benzene versus the extension of the basis functions. The extension 
of the atomic orbitals is determined using the energy shift parameter 
of the SIESTA method~\cite{Artacho-about-Energy-Shift}.
Note that rather extended orbitals are necessary to achieve converged
results. Differences associated with the use of the 
second window
technique introduced in section (\ref{s:two-window-technique}) are of the
order of 0.1~eV. The experimental ionization potential 
is taken from the NIST server \cite{NIST}.
The electron affinity of benzene is taken from \cite{Rienstra-Kiracofe:2001}.
\label{t:ip-ea-benzene}}
\end{table}

The calculations presented in this section needed a fairly
large amount of random access memory (RAM). The amount of RAM
increases as $N^2$ with $N$ the number of dominant products, 
which prohibits the treatment of larger molecules
using the methods described above in a straightforward manner. 
However, as we will see in the next section,
we can use a compression method that dramatically reduces the 
required memory.

\section{Compression of the Coulomb interaction}
\label{s:compression}

As shown in Ref.~\cite{PK+DF+OC}
it is possible to solve the Petersilka-Gossmann-Gross 
equations~\cite{Petersilka} for time-dependent density functional theory (TDDFT)
using a Lanczos type approach if, for example, we are only interested in the
polarizability tensor of the system. In this way, we avoid keeping
the entire linear response matrix $\chi _{\mu \nu }^{0}(\omega )$ in
the computer memory. 
Unfortunately, we were unable to find an analogous Lanczos type procedure
for the self-energy matrix. 
However, we have found an alternative solution to this problem. 
It consists of taking into account the electron dynamics
and keeping preferentially those dominant products that are necessary 
to describe $\chi^{0}_{\mu\nu}$ in the relevant range of frequencies.

\subsection{Defining a subspace within the space of products}

Consider the following  closed form expression of the non interacting response
$\chi_{\mu \nu }^{0}(\omega )$ of eq (\ref{tensorform-response}) 
\begin{equation}
\chi _{\mu \nu }^{0}(\omega +i\varepsilon ) = 2\sum_{E,F}
V _{\mu}^{EF}\frac{n_{F}-n_{E}}{\omega +i\varepsilon -(E-F)}V_{\nu}^{EF}%
\text{, where }V_{\mu }^{EF}=X_{a}^{E}V_{\mu }^{ab}X_{b}^{F}.
\end{equation}
This is a well known expression, but rewritten in the basis of dominant products \cite{DF+PK}.
It must be emphasized that we do not use this equation to compute 
$\chi_{\mu \nu }^{0}(\omega)$ 
(it would require $O(N^4)$ operations) but this explicit representation 
of the exact non interacting response is nonetheless crucial 
for motivating our method of compression.  

Clearly $\chi_{\mu \nu }^{0}(\omega )$ is built up from $O(N^2)$ vectors $V_{\mu }^{EF}$.
On the other hand, the entire space of products is, by construction, of only $O(N)$ dimensions.
Therefore, there must be a significant amount of collinearity in the
set of vectors $V_{\mu}^{EF}$ and a much smaller subset of such
vectors should span the space where $\chi_{\mu\nu}^{0}(\omega)$
acts. As candidates for the generators of this subspace, we 
sort the vectors $\{V_{\mu }^{EF}$ $\}$ according to $|E-F|$
up to a certain \textit{rank} $N_{\mathrm{rank}}$
\begin{equation}
\{X_{\mu }^{n}\}\equiv \text{subset of }\{V_{\mu }^{EF}\}
\text{ limited according to } |E-F|<E_{\mathrm{threshold}},
n=1\ldots N_{\mathrm{rank}}.
\label{subsetof_vef}
\end{equation}%
Here we treat $\{E,F\}$ as electron-hole pairs, i.e. $E<0$ and $F>0$. 

As a first test of whether the subspace carries
enough information, we define a projector onto it
\begin{align}
g^{mn} &=X_{\mu }^{m} v^{\mu\nu} X_{\nu }^{n}; \label{projector0} \\
P_{\mu \nu } &=X_{\mu }^m g_{mn} X_{\nu }^{n}, \text{ where }
      g_{mn}=\left( g^{mn}\right) ^{-1}; \notag \\
P_{\nu }^{\mu } &=v^{\mu\mu'}P_{\mu' \nu }.  \notag 
\end{align}%
It can be shown without difficulty that $P_{\nu }^{\mu }$ is indeed
a projector in the sense of $P^{2}=P$. We can use it to project
the screened Coulomb interaction onto the subspace generated by
the set $\{X_{\mu}^{n}\}_{n=1..N_{\mathrm{rank}}}$ 
\begin{equation}
W_{\mathrm{projected}}^{\mu \nu }(\omega )=
P_{\mu'}^{\mu }P_{\nu'}^{\nu }W^{\mu'\nu'}(\omega).
\label{proj-scr-inter}
\end{equation}%
We must choose $N_{\mathrm{rank}}$ large enough so that the
trace of the projected spectral density 
$W_{\mathrm{projected}}^{\mu \nu }(\omega )$ is sufficiently close
to the original one. We checked that this works even for 
$N_{\mathrm{rank}}$ considerably
smaller than the original dimension of the space of products.
We can go further and reduce the dimension of the subspace by
eliminating collinear vectors from it. We do this by diagonalizing
the matrix $g^{mn}$ in eq (\ref{projector0}) and by defining
new vectors $Z_{\mu}^{\lambda }$\cite{note1}

\begin{align}
g^{mn}\xi_{n}^{\lambda } &=\lambda \xi _{m}^{\lambda }, \notag \\
Z_{\mu}^{\lambda } & \equiv X_{\mu }^{m}\xi _{m}^{\lambda }/\sqrt{\lambda}. 
\label{subspace}
\end{align}%

To define the vector space $\{Z_{\mu}^{\lambda }\}$, we first
discard the  eigenvectors $\xi _{m}^{\lambda }$ that correspond
to eigenvalues $\lambda$ smaller than a threshold with respect to the Coulomb
metric $v^{\mu\nu}$ and we then normalize the remaining vectors, for simplicity.
As a result of this procedure, we obtain a smaller set of vectors that we
denote again by $\{Z_{n \mu}\}$ with 
$n=1\ldots N_{\mathrm{subrank}}$, 
with $N_{\mathrm{subrank}}\leq N_{\mathrm{rank}}$
and the additional property that they are orthonormal with
respect to the Coulomb metric $v^{\mu\nu}$
\begin{equation}
Z_{m}^{\mu }Z_{n \mu}=\delta _{mn}\text{, where }Z_{m}^{\mu }=
v^{\mu\nu }Z_{m \nu}, \text{ for } m,n=1\ldots N_{\mathrm{subrank}}.
\label{global-basis-orthogonality}
\end{equation}%

\subsection{ Construction of the screened interaction
from the action of the response function in the subspace}

From the preceding discussion we know that $\chi _{\mu \nu }^{0}$ can be adequately
represented in the previously constructed subspace $\{Z_{\mu}^{n}\}$
in the sense of 
\begin{equation}
\chi _{\mu \nu }^{0}=Z_{m \mu }\chi _{mn}^{0}Z_{n \nu },\text{ with }%
\chi _{mn}^{0}=Z_{m}^{\mu }\chi _{\mu \nu }^{0}Z_{n}^{\nu }.
\label{subspace_chi0}
\end{equation}%
To see which form the screened Coulomb interaction (\ref{Coulombscreening})
takes for such a density response $\chi_{\mu \nu }^{0}$, we write 
it as a series 
\begin{equation}
W^{\mu \nu }=\left( \frac{1}{v^{-1}-\chi^0}\right)^{\mu \nu }=v^{\mu \nu}
+v^{\mu \mu '}\chi_{\mu '\nu '}^{0} v^{\nu'\nu }+v^{\mu \mu '}
\chi _{\mu'\nu'}^0 v^{\nu'\mu''}\chi_{\mu'' \nu''}^0 v^{\nu''\nu}+\cdots 
\label{W-series}
\end{equation}%
Because $\chi ^{0}$ acts --- by hypothesis --- only in the subspace,
the series may be simplified. Lets insert the representation of
the response function $\chi_{\mu \nu }^{0}$ of eq (\ref{subspace_chi0})
into the series (\ref{W-series}) 
\begin{equation}
W^{\mu \nu }=v^{\mu \nu }+v^{\mu \mu '}
\left[ Z_{m \mu'}\chi _{mn}^{0}Z_{n \nu'}\right]v^{\nu '\nu}+
v^{\mu \mu '}\left[ Z_{m \mu'}\chi_{mn}^{0}Z_{n \nu'}\right]v^{\nu '\mu''}
\left[ Z_{m \mu''}\chi _{mn}^{0}Z_{n \nu''}\right] v^{\nu''\nu}+\cdots,\notag
\end{equation}%
then use the orthogonality property of the basis vectors
$Z_{m \mu}$ (\ref{global-basis-orthogonality}) and find
\begin{align}
W^{\mu \nu } &=v^{\mu \nu }+Z_{m}^{\mu}\chi _{mr}^{0}
\left[ \delta_{rn}+\chi_{rn}^0
+\chi _{rs}^0\chi_{sn}^0+\cdots\right] Z_{n}^{\nu } = \label{chi_RPA} \\
&=v^{\mu \nu }+Z_{m}^{\mu }\chi _{mn}^{\mathrm{RPA}}Z_{n}^{\nu }.  \notag
\end{align}%
Here we introduced the new response function
$\chi_{mn}^{\mathrm{RPA}} \equiv \left( \delta_{mk}-\chi_{mk}^{0}\right)^{-1}\chi_{kn}^{0}$.
From the preceding arguments we conclude that the dynamically
screened Coulomb interaction $W^{\mu \nu }$
can be computed in terms of the response function $\chi_{mn}^{\mathrm{RPA}}$
within the previously constructed subspace and the matrix inversion in this
smaller space is of course much cheaper than in the original space.
This is a welcome feature --- the number of operations for matrix inversion
 scales with the cube of the dimension and
a compression by a factor 10 will lead to a 1000 fold acceleration
of this part of the computation.

It is important to note that, although an energy cut-off $E_{\mathrm{threshold}}$
is used to choose the relevant $\{V_{\mu}^{EF}\}$ vectors,
high frequencies components of the
response and the screened interaction are explicitly 
calculated. 
$E_{\mathrm{threshold}}$ only
serves to construct the frequency independent basis vectors $\{Z_{n}^{\nu}\}$
according to eq (\ref{subspace}). This basis is later used 
to calculate the response function $\chi^0_{\mu\nu}(\omega)$
in the whole frequency range (see eq (\ref{subspace_chi0})).
Of course, we can expect that, if $E_{\mathrm{threshold}}$ 
is chosen too small, the ability of the compressed
basis to represent 
the high energy components of the response 
will eventually deteriorate. However,
we are interested in the low energy excitations of the system
and, as we will show in the next subsection, those can be
accurately described using values of $E_{\mathrm{threshold}}$
that allow for a large reduction in the size of the product
basis. Furthermore, it is also important to note that
the instantaneous self-energy $\Sigma_x$, for which 
a compression criterion based on our definition of $E_{\mathrm{threshold}}$ 
is dubious, is calculated within the original dominant product basis,
i.e. before this non local compression is performed.

\subsection{The compression in the case of benzene }
\label{ss:compression-benzene}

The non local compression depends on two parameters: 
i) the maximum energy $E_{\text{threshold}}$
of the Kohn-Sham electron-hole pairs
 $\{V_{\mu}^{EF}\}$ in eq (\ref{subsetof_vef}) , 
and ii) the eigenvalue threshold $\lambda$ for identifying 
the important basis vectors
$\{Z_{n}^{\nu}\}$ in eq (\ref{subspace}). 

Table \ref{t:ea-benzene}
shows the electron affinity of benzene 
as a function of $E_{\text{threshold}}$ and $\lambda$.
The computational parameters have been chosen as in section \ref{s:results-1}
and the energy shift to define the extension of the orbitals is 3~meV.
Table \ref{t:ea-benzene} illustrates a general feature that we have found in
many test for several systems: 
$N_{\text{rank}}$ can be chosen of the order of the number of 
atomic orbitals $N_{\text{orb}}$. We have found that 
$N_{\text{rank}}\approx 5 N_{\text{orb}}$
usually guarantees a converged result for the HOMO and LUMO 
levels. In any case, the number of 
relevant linear combinations $N_{\text{subrank}}$ was always 
much smaller than the number of dominant 
functions, with a typical compression ratio of ten or more. 

\begin{table}
\centerline{
\begin{tabular}{|c|c|c|c|}
\hline
                               & $\lambda=10^{-2}$ & $\lambda=10^{-3}$ & $\lambda=10^{-4}$ \\
\hline
$E_{\text{threshold}}=10$ eV   &  2.48 (33)  &  2.46 (37)  &  2.45 (39)  \\
$E_{\text{threshold}}=20$ eV   &  1.38 (96)  &  1.39 (133) &  1.40 (171) \\
$E_{\text{threshold}}=40$ eV   &  1.41 (132) &  1.41 (192) &  1.41 (279) \\
\hline
\end{tabular}}
\caption{
Electron affinities for benzene versus the compression parameters $E_{\text{threshold}}$ and
eigenvalues cutoff $\lambda$ in eq (\ref{subspace}). In brackets the dimension of compressed subspace
is given $N_{\text{subrank}}$. The dimension $N_{\text{rank}}$ is governed by $E_{\text{threshold}}$
and was $39$, $297$ and $765$ for $E_{\text{threshold}}=10$ eV, $E_{\text{threshold}}=20$ eV and
$E_{\text{threshold}}=40$ eV, accordingly. Number of atomic orbitals $N_{\text{orb}}=108$, while
the number dominant products is $N_{\text{prod}}=2325$.
\label{t:ea-benzene}}
\end{table}

\begin{figure}[htb]
\begin{tabular}{m{7cm}m{0.1cm}m{5cm}}
\includegraphics[width=7cm,viewport=50 50 410 300, angle=0,clip]{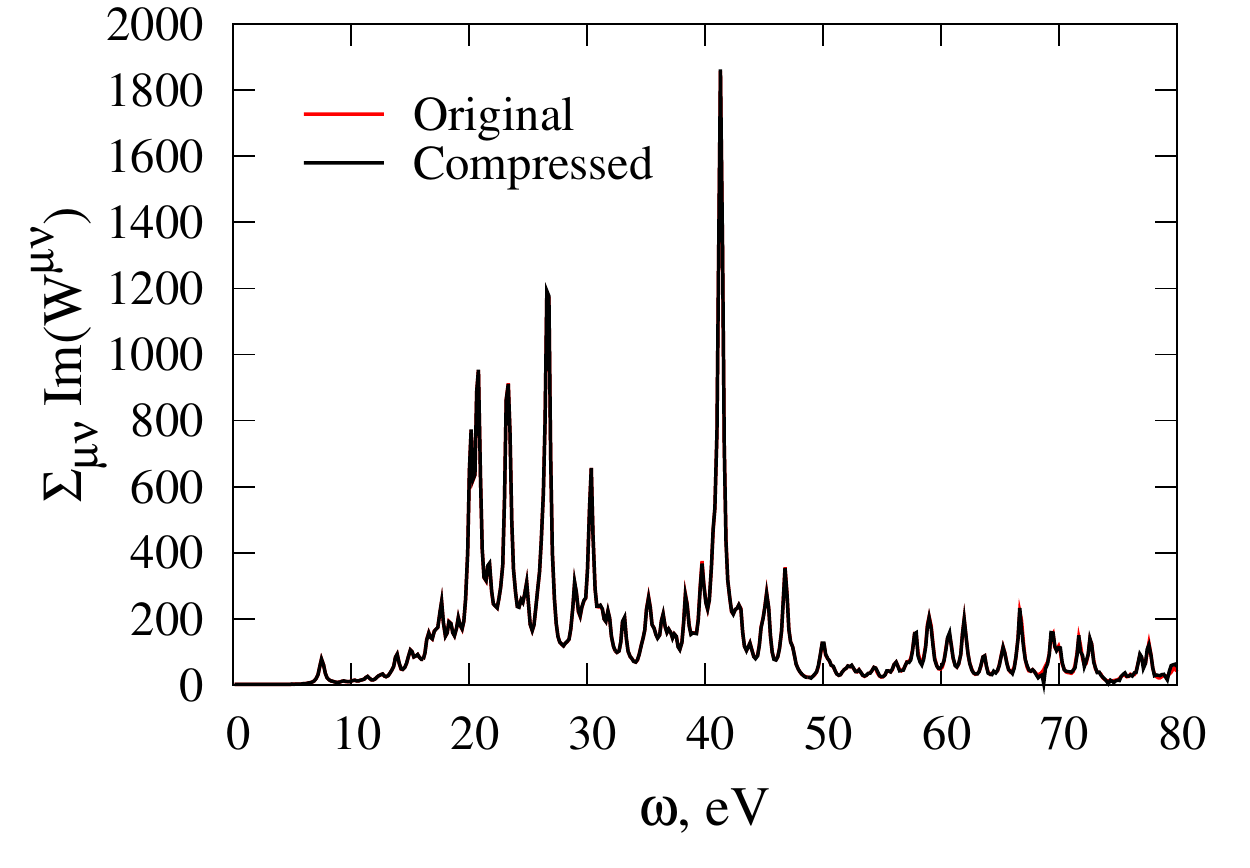} &&
\includegraphics[width=7cm,viewport=50 50 410 300, angle=0,clip]{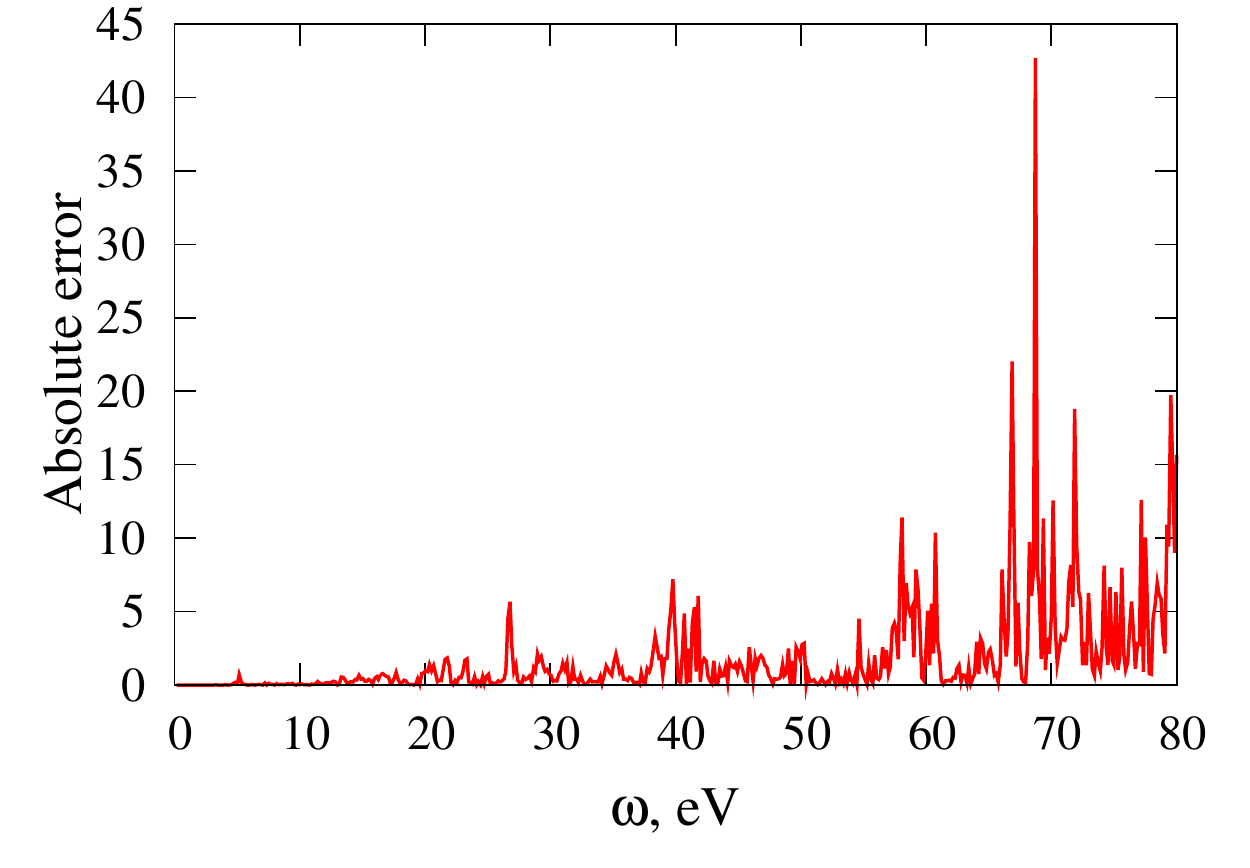} \\
\centerline{a)} && \centerline{b)}
\end{tabular}
\caption{
a) Comparison of the screened interaction calculated 
for benzene using our original dominant product basis and 
the screened interaction projected to a compressed product basis
(see eq~(\ref{proj-scr-inter})). We plot the sum of all the 
matrix elements of the imaginary part of the screened
interaction. 
b) A plot of the difference of the functions represented in panel a. 
The change in spectral weight of  
the screened interaction due to compressing the space 
of dominant products is seen to be small. Please notice
the different scales of the y-axis in both panels.
\label{f:sf-scr-inter-benzene}}
\end{figure}

To further illustrate the quality of the basis, we will compare the trace
of the original screened interaction with the trace of the projected screened
interaction for the benzene molecule. The result of the comparison can be seen
in figure \ref{f:sf-scr-inter-benzene}.
In this test calculation, the dominant product basis consists of 921 functions,
while the compressed basis contains only 248 functions. Examples of compression for
larger molecules will be presented in section \ref{s:results-2}.

The examples presented in this section show that the screened Coulomb interaction
can be effectively compressed. A practical algorithm that uses the non local compression 
and maintains the $O(N^3)$ complexity scaling of the calculation will be presented
in the next section.

\section{Maintaining $O(N^3)$ complexity scaling by compressing / decompressing}
\label{s:maintain-n3}

The favorable $O(N^{2})$ scaling of the construction of the uncompressed
non interacting response $\chi^{0}_{\mu\nu}$ 
is due to its locality. On the other hand,
we need compression for $\chi^{0}_{\mu\nu}$ to fit into the computer memory and
the compressed $\chi^{0}_{mn}$ is no longer local. To satisfy the
two mutually antagonistic criteria of 
\textit{i}) locality (for computational speed) and 
\textit{ii}) small dimension (to fit into the computer memory) we shuttle back and
forth as needed between the uncompressed/local and the compressed/non local
representations of the response $\chi^{0}$ and of the screened interaction $W$. 
Both compression and decompression are matrix operations that scale as
$O(N^{3})$ and this, along with the matrix inversion in eq (\ref{chi_RPA})
in the computation of the screened Coulomb interaction, and the computation
of the spectral densities $\rho_{ab}^{\pm}(s)$ is the reason why our 
implementation
of $GW$ scales as $O(N^{3})$. 

\subsection{A construction of the subspace response in $O(N^{3})$ operations}

Let us describe an efficient construction of the response $\chi_{\mu \nu}^{0}$ 
and its compressed counterpart $\chi_{mn}^{0}$ 
that, besides, gives us an opportunity to describe our
use of frequency and time domains during the calculation. 
Consider eq (\ref{chi_by_spectra}) that involves convolutions
of the spectral functions $\rho_{bc}^{+}(\omega)$ and
$\overline{\rho}^{-}_{ad}(\omega)\equiv\rho^{-}_{ad}(-\omega)$.  
To make use of the convolution theorem, we will first compute the 
spectral function of the non interacting response $a_{\mu \nu}(s)$ in the time domain
\begin{align}
a_{\mu \nu }(t) &=\int \frac{ds}{2\pi }
a_{\mu \nu }(s)e^{\mathrm{i}st}=2\pi
\int_{0}^{\infty }V _{\mu }^{ab}\rho_{bc}^{+}(s_{1})
e^{\mathrm{i}s_1t}\frac{ds_{1}}{2\pi }\cdot %
\int_{0}^{\infty }V _{\nu }^{cd}\rho
_{ad}^{-}(-s_{2})e^{\mathrm{i}s_2 t }\frac{ds_{2}}{2\pi } \notag \\
&=2\pi V _{\mu }^{ab}\rho _{bc}^{+}(t)V _{\nu }^{cd}
\overline{\rho}_{ad}^{-}(t). \label{sf-of-response-time} 
\end{align}%
In other words, we prepare the use of the FFT driven convolution
by first computing the Fourier transforms of the electronic
spectral densities $\rho^{\pm }$ and once $a_{\mu \nu }(t)$ is computed, we
return to $a_{\mu \nu }(s)$ by an inverse Fourier transformation.
This is nothing else but the fast convolution method with the
Fourier transform of the spectral densities $\rho_{bc}^{+}(\omega)$ and
$\overline{\rho}_{ad}^{-}(\omega)$ carried out prior to the tensor operations in 
eq (\ref{sf-of-response-time}).

Above we saw how to compute the spectral function of the non interacting density response.
However, as we mentioned before,
we cannot easily store this information in the memory of the computer and we must therefore compress
this quantity as soon as it is found to avoid over flooding the computer memory. An efficient way to do this 
is to compute $a_{\mu \nu }(t)$, the spectral function of the non interacting response in the time domain, 
in a ``time by time'' fashion, with the time variable $t$ 
in the outer loop. Fortunately, the compression of the
response $\chi^0_{\mu\nu}$ according to equation (\ref{subspace_chi0})
can be done on the level of its spectral function $a_{\mu \nu }(t)$
separately for each time $t$.

\subsection{A construction of the self-energy in $O(N^{3})$ operations}

Although we use the spectral function given by eq (\ref{spectral_3}) to compute
the self-energy in the second of eqs (\ref{spectral_1}),
it is useful to think also
of  eq (\ref{tensorform-self-energy}) that corresponds to the
Feynman diagram of Figure \ref{f:self-energy-diagram} and 
which has the same locality properties.
Please recall that the product vertex in eq (\ref{VertexDefinition}) is sparse and local
and that the indices $\{a,a',\mu \}$ and $\{b,b',\nu\}$ must each reside on
a single pair of overlapping atoms. Once the indices $a,b$ of the self-energy
are specified, there are only $O(N^{0})$ possibilities of choosing the remaining
indices.
Therefore, the calculation of $\Sigma ^{ab}(t)$ requires
asymptotically $O(N^{2})$ operations 
provided that the screened Coulomb interaction $W^{\mu\nu}$
in a basis of localized functions is known.
However, the local screened Coulomb interaction  $W^{\mu \nu }$ in the original
space of dominant products does not fit into the computer memory as opposed
to the compressed, but non local, response $\chi_{mn}^{\mathrm{RPA}}$ that we store
(see eq (\ref{chi_RPA})).
We may, however, regain locality by decompressing $\chi_{mn}^{\mathrm{RPA}}$
at the cost of $O(N^{3})$ operations, using the identity 
$W_{\mathrm{dynamical}}^{\mu \nu }=Z_{m}^{\mu}\chi_{mn}^{\mathrm{RPA}}Z_{n}^{\nu}$
in eq (\ref{chi_RPA}). As we cannot keep $W_{\mathrm{dynamical}}^{\mu \nu }$ in
the computer memory, we must try to ``decompress $\chi_{mn}^{\mathrm{RPA}}$ on the fly''.
To do this, let us transform the first of eqs (\ref{spectral_3}) into
the time domain. 
For instance, for the positive part of the spectral density
$\sigma_{+}^{ab}(t)$ of the self-energy, we find

\begin{equation}
\sigma_{+}^{ab}(t)=2\pi V_{\mu }^{aa'}\rho _{a'b'}^{+}(t)
V_{\nu }^{b'b}\gamma_{+}^{\mu \nu }(t).  \notag
\end{equation}%
Again, the representation in time of $\rho_{a'b'}^{+}(t)$ is prepared only once.
However, the transform 
$\gamma_{+}^{\mu \nu}(t)=-\frac{1}{\pi }
Z_{m}^{\mu} \mathrm{Im}\chi_{mn}^{\mathrm{RPA}}(t)Z_{n}^{\nu }$
for all times does not fit into the computer memory. 
Therefore we also decompress $\gamma_{+}^{\mu \nu }(t)$ time by time by letting
the time $t$ run in the outer loop, by computing $\gamma _{+}^{\mu \nu }(t)$
via decompression for a single time, and by storing only the result
$\sigma _{\pm }^{ab}(t)$ for each time. Once we have computed
$\sigma _{\pm }^{ab}(t)$ for all times, we can find $\sigma_{\pm }^{ab}(s)$
from it.

\section{A summary of the complete algorithm}
\label{s:algorithm-summary}

At this point, it is useful to briefly recapitulate the different
steps of our implementation of Hedin's $GW$ approximation.  
It consists of the following steps:

\begin{enumerate}
\item Export the results of a DFT code that uses numerical
local atomic orbitals as a basis set. 
Here we use the SIESTA code~\cite{siesta},
but other codes like the FHI-AIMS code~\cite{aims} 
could also be used.

\item Set up a basis of dominant products in $O(N)$ operations. Here we 
use the method of Ref.~\cite{DF}.

\item Set up a space of reduced dimension where the 
screened Coulomb interaction will act
and exploiting the low effective rank of this set. Such a subspace is
determined by a set of $N_{\mathrm{rank}}$ vectors 
$V _{\mu }^{EF}$ that correspond to 
electron-hole pairs with a predetermined maximum value of $|E-F|$. 
Further compression is obtained  by diagonalizing the Coulomb metric projected
onto this subspace. This step requires $O(N^3)$ operations. 

\item Choose low and high energy spectral windows and a frequency grid.
Prepare the electronic spectral density $\rho_{ab}(s)$ in these two windows from
the output of the DFT calculation. 

\item Find $\chi_{mn}^{\mathrm{RPA}}$ by constructing and compressing
the local $\chi_{\mu \nu }^{0}$ ``on the fly'' in $O(N^{3})$ operations
and by solving for $\chi_{mn}^{\mathrm{RPA}}$ 
for all frequencies in $O(N^{3})$
operations. The construction must be done in two frequency windows. 
Truncate the spectral
data where needed in order to avoid double counting and store
$\chi_{mn}^{\mathrm{RPA}}$ in the two windows.

\item Find the spectral function of the self-energy by decompressing
$\chi_{mn}^{RPA}\rightarrow W^{\mu \nu }$ ``on the fly'' in $O(N^3)$ operations
and by convolving it with the electronic spectral function. Again this must
be done in two frequency windows and the results must be combined consistently.

\item Construct the self-energy from its spectral representation.

\item Solve Dyson's equation and find the density of states 
from the interacting
Green's function. Obtain the desired spectroscopic information from
the density of states. 
\end{enumerate}

Results obtained with the above algorithm will be discussed in the next section.

\section{Tests for molecules of intermediate size \label{s:results-2}}

The compression technique has been carefully tested in the case
of the benzene molecule. The tests show excellent agreement of the density
of states computed with and without compression. In this section, we will
consider larger molecules such as the 
hydrocarbons naphthalene and anthracene~\cite{comp-details}.
These molecules are well known to differ in their
character as electron acceptors:
naphthalene, like benzene, has a 
negative electron affinity,
while anthracene is an electron acceptor with 
positive electron affinity. 

A compression of the dominant product basis is necessary 
to treat the molecules considered in this section. These molecules 
are too large for a calculation without compression 
on ordinary desktop machines because of memory requirements. 
For naphthalene the dominant product basis contained 4003 functions,
which were reduced to 433 functions after compression.
In the case of anthracene, the dominant product basis contained
5796 functions, while the compressed basis had only 598 functions.

\begin{table}\centerline{
\begin{tabular}{|c|c|c||c|c||c|c|}
\hline
                 & \multicolumn{4}{c||}{Naphthalene} & \multicolumn{2}{c|}{Anthracene} \\
\hline
Energy-shift,    & \multicolumn{2}{c||}{One window}  
                                    & \multicolumn{2}{c||}{Two windows} 
                                                      & \multicolumn{2}{c|}{Two windows}\\
meV              & IP, eV & EA, eV & IP, eV & EA, eV & IP, eV & EA, eV \\
\hline
200              &  7.24  & -0.68   &  7.27  & -0.79  & 6.44   & 0.20  \\
20               &  7.61  & -0.083  &  7.67  & -0.18  & 6.89   & 0.77  \\
\hline
Experiment       &  8.14  & -0.191  &  8.14  & -0.191 & 7.439  & 0.530 \\
\hline
\end{tabular}}
\caption{
Ionization potentials and electron affinities for naphthalene and anthracene
and their dependence on the extension of the atomic orbitals. 
For naphthalene we compared the results obtained with
spectral functions discretized in one or two windows.
The experimental data has been taken from the NIST server~\cite{NIST}.
For naphthalene and anthracene, vertical ionization potentials are
not available at the NIST database. Therefore we give experimental
ionization energies including effects of geometry relaxation.
\label{t:ip-ea-naphthalene-anthracene}}
\end{table}

\begin{figure}[htb]
\centerline{
\begin{tabular}{m{7cm}m{0.1cm}m{5cm}}
\centerline{\includegraphics[width=7cm,viewport=50 50 410 300, angle=0,clip]{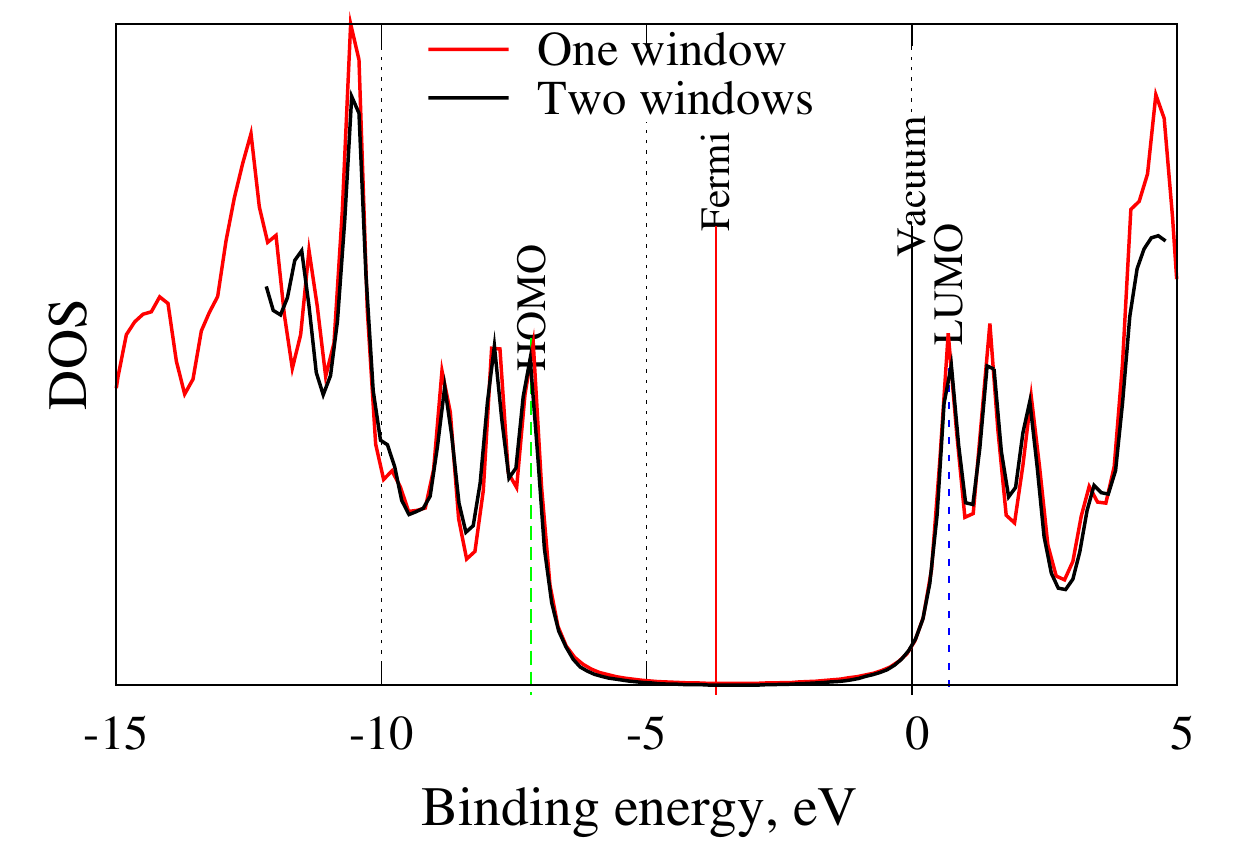}} &&
\centerline{\includegraphics[width=5cm, angle=0,clip]{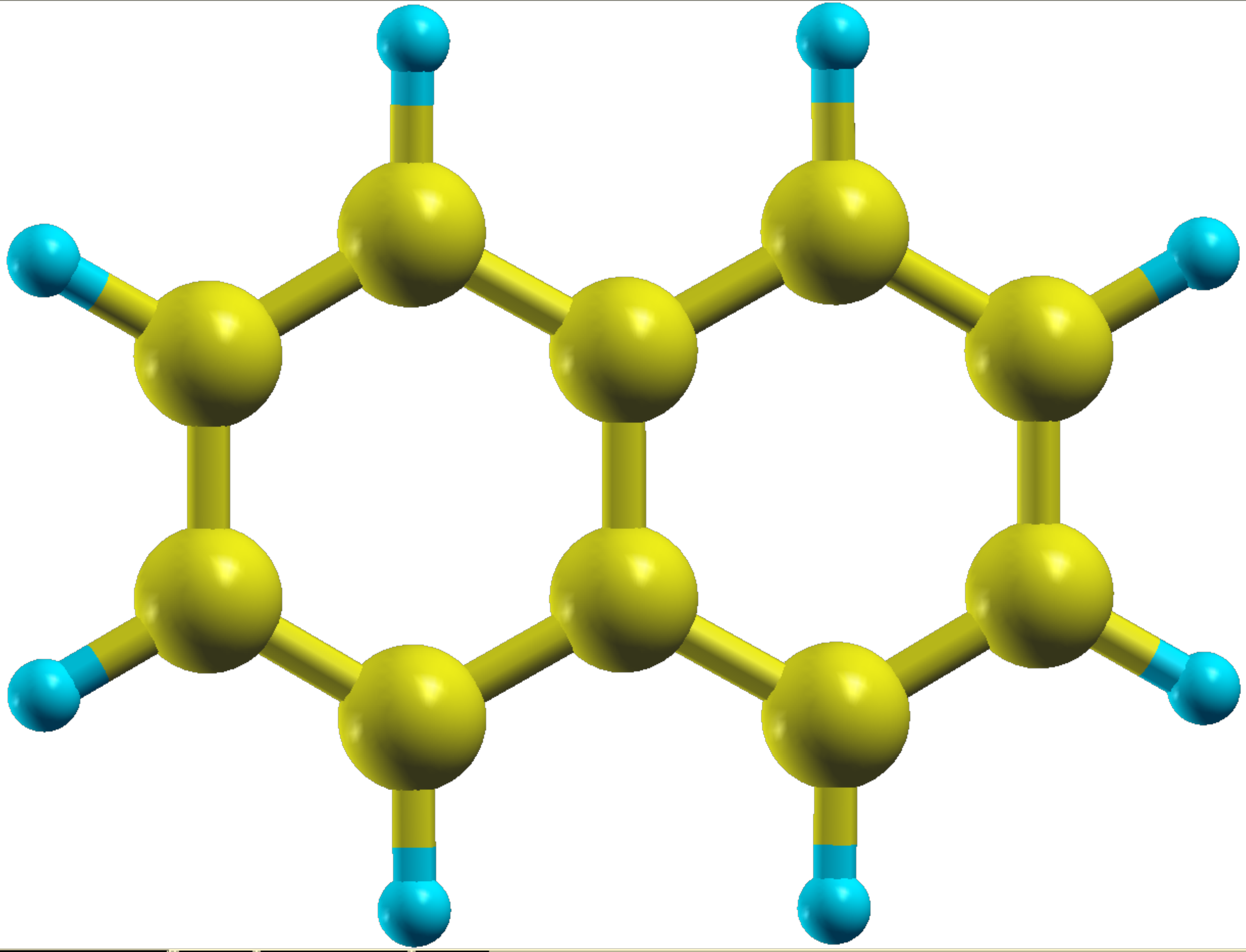}} \\[-0.2cm]
\centerline{a)} && \centerline{b)}
\end{tabular}}
\caption{a) Density of states for naphthalene.
The results
have been obtained with our most extended basis orbitals (corresponding
to an energy shift of 20~meV~\cite{Artacho-about-Energy-Shift}).
We can appreciate the accuracy of the second window technique.
b) Ball and stick model  of naphthalene produced with the 
XCrysDen package \cite{xcrysden}.
\label{f:dos-naphthalene}}
\end{figure}

Table \ref{t:ip-ea-naphthalene-anthracene}
shows our results for naphthalene and anthracene. 
The computational details were similar to those used for benzene 
and already described in section~\ref{s:results-1}.
The two-window results were obtained with frequency grids of only
128 points for naphthalene (in the ranges $\pm$~8.32 eV and $\pm$~80 eV) 
and yet it provides an accuracy on the 0.1 eV level,
while the one-window calculation is done again with
1024 frequencies (in the range of $\pm$~80 eV).
In the case of anthracene, results using frequency grids
of 256 points (in the ranges $\pm$~16 eV and $\pm$~90 eV) are presented.

Again we find a large improvement over the position
of the Kohn-Sham levels in a  DFT-LDA calculation.
The agreement with the experimental data
is certainly improved, although there are still significant
deviations, particularly with respect to the reported ionization
potentials. Interestingly, however, our calculations
recover the important
qualitative feature of anthracene being an electron acceptor.
In the case of anthracene, the results obtained with our most
extended basis orbitals (energy shift of 20~meV) are
in excellent agreement with the recent calculations
by Blase \textit{et al.}~\cite{Blase}. In the case
of naphthalene we can see that the two frequency windows
technique introduces only tiny differences (below 0.1~eV) 
in the positions of the HOMO and LUMO levels.

\begin{figure}[htb]
\centerline{
\begin{tabular}{m{7cm}m{0.1cm}m{7cm}}
\centerline{\includegraphics[width=7cm,viewport=50 50 410 300, angle=0,clip]{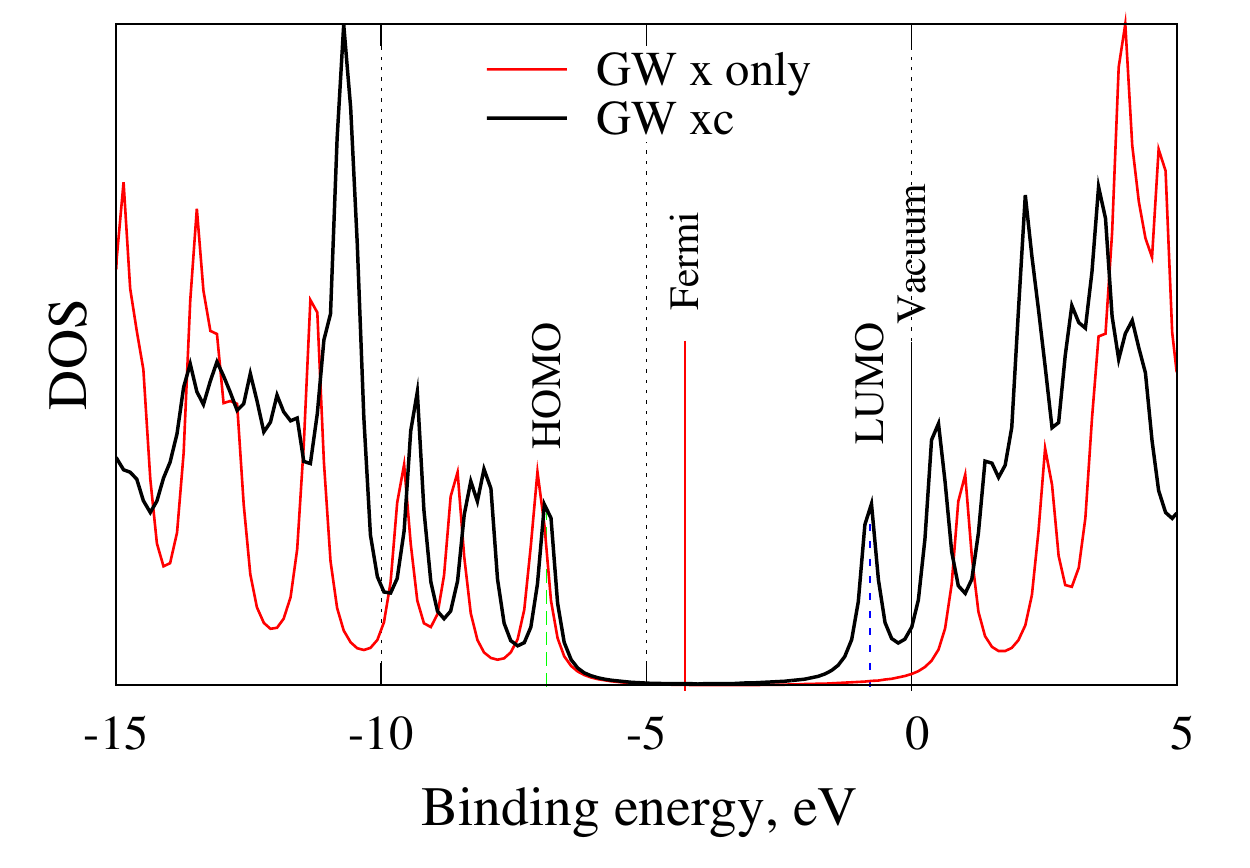}} &&
\centerline{\includegraphics[width=7cm, angle=0,clip]{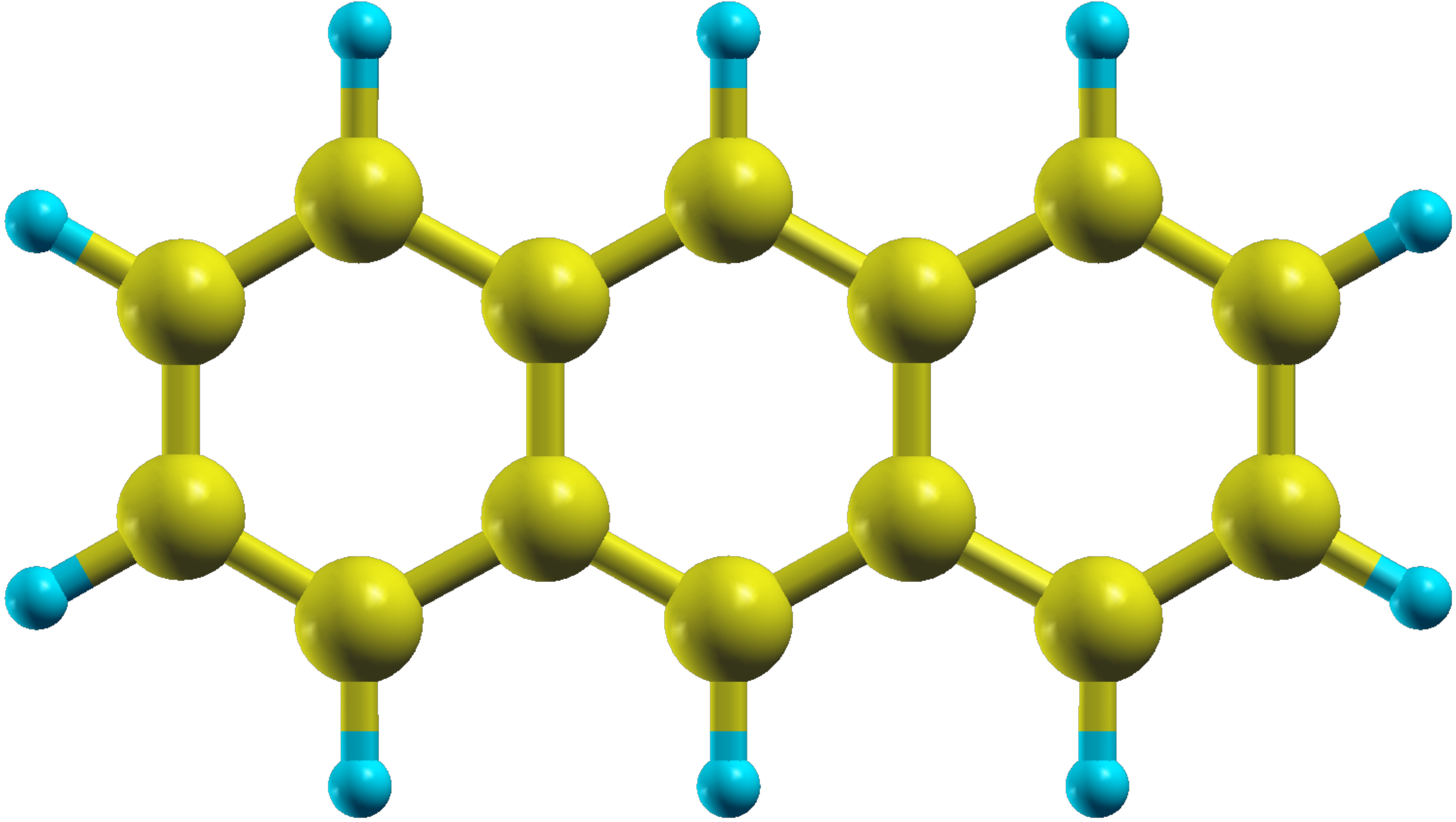}} \\[-0.2cm]
\centerline{a)} && \centerline{b)}
\end{tabular}}
\caption{
\label{f:dos-anthracene}
a) Density of states for anthracene. 
The results have been obtained using  
the extended basis orbitals corresponding
to an energy shift of $20$~meV~\cite{Artacho-about-Energy-Shift}.
Here we compare calculations using
the instantaneous (exchange-only) self-energy and
the full self-energy (including correlation effects). 
The correlation component of the self-energy is crucial 
to reproduce the experimental observation that anthracene
is an acceptor. In contrast, the exchange-only calculation locates
the LUMO level above the vacuum level.
b) Ball and stick model of anthracene produced with the
XCrysDen package \cite{xcrysden}.
}
\end{figure}

The corresponding DOS is shown in 
figures \ref{f:dos-naphthalene} and \ref{f:dos-anthracene}.
One can see in figure \ref{f:dos-anthracene} that 
it is the dynamical part of the self-energy, including
correlation effects, that turns our theoretical 
anthracene into an acceptor, while including only the 
instantaneous self-energy predicts anthracene to be a donor.

These results for molecules of modest size are just a
first application of our algorithm. With its favorable scaling,
our method aims at $GW$ calculations 
for larger molecules of the type used in organic semiconductors. 
However, before carrying out such studies, we should reduce the initial
number of dominant products i.~e. before any compression is applied to it.

\section{Conclusions and outlook}
\label{s:conclusion}

In the present paper we have described our approach to 
Hedin's $GW$ approximation for finite systems. This approach
provides results for densities of states and gaps that are 
in reasonable  agreement with experiment and it requires only
modest computer resources~\cite{comp-details} for the systems
presented here. The complexity of our
algorithm scales asymptotically as the third power of the number of atoms,
while the needed memory grows with the second power of the number of atoms.
We hope that these features, along with a further reduction
of the size of the basis describing the 
products of  localized 
orbitals, will allow to apply our method to describe 
the electronic structure of large molecules 
and contribute to an ab-initio design of organic semiconductors
for technological applications.

The algorithm described here is built upon the LCAO technique~\cite{siesta}
and uses a previously constructed basis in the space of orbital products 
that preserves their locality and avoids fitting procedures~\cite{DF}. 
Moreover, a (non local) compression technique has been developed to
reduce the size of this basis. This allows to store the whole
matrix representation of the screened Coulomb interaction at
all time/frequencies
in random access memory
while significantly reducing the computational time.
The time (and frequency) dependence
of observables is treated with the help of spectral functions. This  avoids
analytical continuations  and allows for operations to be 
accelerated by the use of FFTs.
As a useful byproduct of our focus on spectral functions we obtain, 
as primary result,
an electronic spectral function of the type observed in photo-emission 
and from which we
then read off the HOMO and LUMO levels.

We have applied our method to benzene, naphthalene and anthracene.
As expected, we find that our estimations of the HOMO and LUMO positions
and the corresponding gaps are significantly improved over the results
obtained from the Kohn-Sham eigenvalues in a plain DFT-LDA calculation.
Our results approach the experimental data but, as observed
by other authors~\cite{Blase}, these ``single-shot'' $GW$-LDA calculations
(or $G_0W_0$-LDA using a more 
standard terminology) 
still present sizeable deviations from the measured ionization
potentials and electron affinities. 
In  general, our results are in good agreement with previous      
$G_0W_0$-LDA calculations for similar
systems~\cite{Blase,Baroni$GW$,Tiago-Chelikowsky-and-Frauenheim}.
Thus, we expect further improvements
by iterating our procedure until self consistency or, as suggested
by other authors in the case relatively small
molecules~\cite{Hahn2005,Blase,DanishWannierMolecules}, 
by using Hartree-Fock results as an input for our $G_0W_0$ calculations.
For periodic systems it is well known that $G_0W_0$-LDA systematically
underestimates the size of the gaps of semiconductors. The best results
so far were found using the so-called ``improved quasi-particle method'' 
\cite{Schilfgaarde-Kotani-Faleev:2006, chinese-impl-of-kotani}.
A realization of this method in our framework should also improve 
the precision of our results.

The method presented in this paper depends crucially 
on the quality and size of the original LCAO basis.
A possible limitation is that 
the typical LCAO basis used in electronic structure calculations 
are constructed and optimized in order to describe ground-state 
properties~\cite{Artacho-about-Energy-Shift}. However,
it is possible to optimize an LCAO basis, for example
using a technique similar to that described in Ref.~\cite{Daniel},
to represent electronically excited states. This will increase
the accuracy and applicability of the method and could even 
allow to reduce the size of the original LCAO basis used to 
represent the electronic states.
Moreover, by comparing our basis with that of other authors, 
there are indications that the (local) basis
of dominant products used in this paper
can be reduced in dimension without changing 
the physical results~\cite{Blase}. 
Such a reduction should lead to an important improvement 
of the prefactor in our implementation
of $GW$, but also, as a side effect, introduce a similar
acceleration in our published TDDFT algorithm \cite{PK+DF+OC} that is already
competitive, in its present form, with other TDDFT codes.

The quantities calculated in the presented algorithm can be
useful in other branches of many-body perturbation theory.
For instance, the screened Coulomb interaction is
a crucial ingredient of the Bethe-Salpeter technique that
is needed to study excitons and the optical response of excitonic systems.
In this context it is interesting to note 
\cite{Benedict} that the solution of the Bethe-Salpeter equation scales
as $O(N^{3})$ for clusters of size $N$, at least when suppressing 
the dynamic part of
the fermion self-energy and the dynamic part of the screening of
the Coulomb interaction.  Calculations of 
the transport properties of  molecular junctions~\cite{Brandbyge2002}
are another possible application of the $GW$ approach described here.

\section*{Acknowledgments}

We thank Olivier Coulaud for useful advice on computing, and Isabelle Baraille and Ross Brown 
for discussions on chemistry, both in the context of the ANR project ``NOSSI''.
James Talman has kindly provided essential computer algorithms and codes, 
and we thank him, furthermore, for inspiring discussions and 
for correspondence. We are indebted to the organizers 
of the ETSF2010 meeting at Berlin
for feedback and perspective on the ideas of this paper.  
Arno Schindlmayr, Xavier Blase and Michael Rohlfing helped with extensive
correspondence on various aspects of the $GW$ method.
DSP and PK acknowledge financial support from 
the Consejo Superior de Investigaciones Cient\'{\i}ficas (CSIC),
the Basque Departamento de Educaci\'on, UPV/EHU (Grant No. IT-366-07), 
the Spanish Ministerio de Ciencia e Innovaci\'on 
(Grants No. FIS2007-6671-C02-02 and FIS2010-19609-C02-02) and,
the ETORTEK program funded by the Basque Departamento de Industria and the
Diputaci\'on Foral de Guipuzcoa.


\begin{thebibliography}{99}

\bibitem{Hedin} L.~Hedin, Phys. Rev. \textbf{139}, A796 (1965).
For a review see F.~Aryasetiawan and O.~Gunnarsson,
Rep. Prog. Phys. \textbf{61}, 237 (1998); 
C.~Friedrich and A.~Schindlmayr, \textit{Many-Body Perturbation Theory: The $GW$ Approximation},
NIC Series, \textbf{31}, 335 (2006)
\url{http://www.fz-juelich.de/nic-series/volume31/friedrich.pdf}

\bibitem{Schilfgaarde-Kotani-Faleev:2006}
M.~van~Schilfgaarde, T.~Kotani, S.~Faleev,
% Quasiparticle Self-Consisten $GW$ Theory
Phys. Rev. Lett. \textbf{96}, 226402 (2006); especially 
figure 1 of this paper.

\bibitem{LouieRohlfing} M.~Rohlfing and S.~G.~Louie, 
Phys. Rev. B \textbf{62}, 4927 (2000).

\bibitem{OrganicsReview} H.~Hoppe, N.~S. Sariciftci, J. Mater. Res.
\textbf{19}, 1924 (2004).

\bibitem{Blase} 
X.~Blase, C.~Attaccalite, V.~Olevano, 
%First-principles $GW$ calculations for fullerenes, porphyrins, phtalocyanine,
% and other molecules of interest for organic photovoltaic applications
Phys. Rev. B \textbf{83}, 115103 (2011).

\bibitem{RohlfingRetinol} M.~S.~Kaczmarski, Y.~Ma, and M.~Rohlfing,
Phys. Rev. B \textbf{81}, 115433 (2010).

\bibitem{private-comm-scaling}X.~Blase, private communication (15 December 2010);
J.~Chelikowsky and M.~Tiago, private communication (15 December 2010).

\bibitem{Furche} H.~Eshuis, J.~Yarkony, and F.~Furche, J. Chem. Phys. 
\textbf{132}, 234114 (2010).

\bibitem{Godby} M.~M.~Rieger, L.~Steinbeck, I.~D.~White, H.~N.~Rojas,
R.~W.~Godby, Comp. Phys. Comm. \textbf{117}, 211 (1999).

\bibitem{siesta} J.~M.~Soler, E.~Artacho, J.~D.~Gale, A.~Garc\'{\i}a,
J.~Junquera, P.~Ordej\'{o}n and D.~S\'{a}nchez-Portal,
J. Phys.: Condens. Matter \textbf{14}, 2745 (2002);
E.~Artacho, E.~Anglada, O.~Dieguez, J.~D.~Gale, A.~Garc\'{\i}a,
J.~Junquera, R.~M.~Martin, P.~Ordej\'{o}n, J.~M.~Pruneda,
D.~S\'{a}nchez-Portal and J.~M.~Soler, J. Phys.: Condens. Matter 
\textbf{20}, 064208 (2008).

\bibitem{reviewPayne} M.~C.~Payne, M.~P.~Teter, D.~C.~Allan, T.~A.~Arias
and J.~D.~Joannopoulos, Rev. Mod. Phys. \textbf{64}, 1045 (1992)

\bibitem{DF} D.~Foerster, J. Chem. Phys. \textbf{128}, 34108 (2008).

\bibitem{DF+PK} D.~Foerster and P.~Koval,
J. Chem. Phys. \textbf{131}, 044103 (2009).

\bibitem{PK+DF+OC} P.~Koval, D.~Foerster and O.~Coulaud, 
J. Chem. Theory Comput. \textbf{6}, 2654 (2010).

\bibitem{Licence} The code solves the Petersilka-Gossmann-Gross equations of
TDDFT linear response and will become available under an appropriate non commercial license.

\bibitem{RealSpaceChelikowsky} A.~Natan, A.~Benjamini, D.~Naveh, L.~Kronik,
M.~L.~Tiago, S.~P.~Beckman, and J.~R.~Chelikowsky, Phys. Rev. B \textbf{78},
75109 (2008).

\bibitem{BigDFT} L.~Genovese, A.~Neelov, S.~Goedecker, T.~Deutsch,
S.~A.~Ghasemi, A.~Willand, D.~Caliste, O.~Zilberberg, M.~Rayson,
A.~Bergman and R.~Schneider, J. Chem. Phys. \textbf{129}, 014109 (2008).

\bibitem{Wannier} N.~Marzari and D.~Vanderbilt,
% Maximally localized generalized Wannier functions for composite energy bands
Phys. Rev. B \textbf{56}, 12847 (1997).

\bibitem{Baroni$GW$} P. Umari, G. Stenuit and S. Baroni,
Phys. Rev. B \textbf{79}, 201104R (2009); 
Phys. Rev. B \textbf{81}, 115104 (2009).

\bibitem{DanishWannierMolecules} C.~Rostgaard, K.~W.~Jacobsen, and
K.~S.~Thygesen, Phys. Rev. B \textbf{81}, 085103 (2010).

\bibitem{RPA} D.~Pines, \textit{Elementary Excitations in Solids} 
(Wiley, New York, 1964).

\bibitem{Debyemodel} P. M. Chaikin and T. C. Lubensky,
\textit{Principles of Condensed Matter Physics} (Cambridge University Press, 1995). 

\bibitem{ManyBodyText} A.~L.~Fetter and J.~D.~Walecka,
\textit{Quantum Theory of Many-Particle Systems} (Dover, New York, 2003).

\bibitem{HK} P. Hohenberg and W. Kohn, Phys. Rev. \textbf{136}, B864 (1964).

\bibitem{KS} W. Kohn and L. J. Sham, Phys. Rev. \textbf{140}, A1133 (1965). 

\bibitem{PhotoEmissionExperiments} F.~Reinert, S.~H\"{u}fner,
New J. Phys. \textbf{7}, 97 (2005).

\bibitem{Fulde} P.~Fulde, \textit{Electron Correlations in Molecules and Solids},
Vol. \textbf{100}, in Springer Series in Solid-State Sciences
(Springer, Berlin, 1991).

\bibitem{Martin} R.~M.~Martin, \textit{Electronic Structure: Basic Theory and
Practical Methods} (Cambridge University Press, Cambridge, 2004).

\bibitem{LinearDependenceOldpaper} J.~E.~Harriman,
Phys. Rev. A \textbf{34}, 29 (1986).

\bibitem{Talman} J.~D.~Talman,
J. Chem. Phys. \textbf{80}, 1984 (2000);
J. Comput. Phys. \textbf{29}, 35 (1978);
Comput. Phys. Commun. \textbf{30}, 93 (1983);
Comput. Phys. Commun. \textbf{180}, 332 (2009).

\bibitem{NumericalRecipesFFT} W.~H.~Press, S.~A.~Teukolsky,
W.~T.~Vetterling, and B.~P.~Flannery, \textit{Numerical Recipes}
(Cambridge University Press, Cambridge, 2007).



\bibitem{Shishkin-Kresse:2006}
The technique of discretizing spectral functions was also used
by M.~Shishkin and G.~Kresse, 
%Implementation and performance of the frequency-dependent $GW$ method
%within the PAW framework
Phys. Rev. B \textbf{74}, 035101 (2006).

\bibitem{Berger}
P.~Umari, G.~Stenuit and S.~Baroni, Phys. Rev. B \textbf{81}, 115104 (2010);
J.~A.~Berger, L.~Reining, and F.~Sottile, Phys. Rev. B \textbf{82}, 41103R (2010).

\bibitem{comp-details} All the calculations presented in this
paper have been performed on one core of an 
Intel Core Quad CPU Q9400 \@ 2.66 GHz using the MKL BLAS library.

\bibitem{Artacho-about-Energy-Shift} J.~Junquera, \'O.~Paz, D.~S\'anchez-Portal, and E.~Artacho,
Phys. Rev. B \textbf{64}, 235111 (2001).

\bibitem{Perdew-Zunger} J. P. Perdew and A. Zunger, 
Phys. Rev. B \textbf{23}, 5048 (1981)

\bibitem{Troullier-Martins} N. Troullier and J. L. Martins, 
Phys. Rev. B \textbf{43}, 1993 (1991)

\bibitem{Tiago-Chelikowsky-and-Frauenheim}
M.~L.~Tiago and J.~R.~Chelikowsky,
%Optical excitations in organic molecules, clusters and defects studied by 
% first-principles Green's function methods
Phys. Rev. B \textbf{73}, 205334 (2006);
T.~A.~Niehaus, M.~Rohlfing, F.~Della~Sala, A.~Di~Carlo, and Th.~Frauenheim,
% Quasiparticle energies for large molecules: a tight-binding-based GF approach
Phys. Rev. A \textbf{71}, 022508 (2005).

\bibitem{xcrysden}A.~Kokalj, Comp.~Mater.~Sci. \textbf{28}, 155 (2003).

\bibitem{NIST}\url{http://cccbdb.nist.gov/}

\bibitem{Rienstra-Kiracofe:2001} J.~C.~Rienstra-Kiracofe, Ch.~J.~Barden, Sh.~T.~Brown,
and H.~F.~Schaefer, J. Phys. Chem. A \textbf{105}, 524 (2001).
 
\bibitem{Petersilka} M.~Petersilka, U.~J.~Gossmann, and E.~K.~U.~Gross,
Phys. Rev. Lett.\textbf{76}, 1212 (1996);
M.~E.~Casida, in \textit{Recent Advances in Density Functional Theory},
edited by D.~P.~Chong (World Scientific, Singapore, 1995, p.~155).

\bibitem{note1}
Including $O(N^2)$ electron-hole pairs $EF$ into the diagonalization procedure would
scale as $O(N^6)$, but this is unnecessary, as the total dimension of the product space
is only $O(N)$. Furthermore, our computations show explicitly that a small number of
low energy pairs is already sufficient to get correct results
(see subsection \ref{ss:compression-benzene}).

\bibitem{aims}V.~Blum, R.~Gehrke, F.~Hanke, P.~Havu, V.~Havu, X.~Ren, K.~Reuter,
and M.~Scheffler, Comput. Phys. Commun. \textbf{180}, 2175 (2009).

\bibitem{Hahn2005} P. H. Hahn, W. G. Schmidt and F. Bechstedt,
Phys. Rev. B \textbf{72}, 245425 (2005).

\bibitem{chinese-impl-of-kotani}
San-Huang Ke, 
% All-electron $GW$ calculation for molecules: ionization energy and electron affinity of conjugated molecules
eprint arXiv: 1012.1084

\bibitem{Daniel} D.~S\'anchez-Portal, E.~Artacho and J.~M.~Soler,
J. Phys.: Condens. Matter \textbf{8}, 3859 (1996).
%There is ongoing work by Lange and Neugebauer on the topic.
 
\bibitem{Benedict}
L.~X.~Benedict and E.~L.~Shirley, 
%Ab initio calculation of dielectric function including the electron-hole interaction:
%Application to GaN and CaF2
Phys. Rev. B \textbf{59}, 5441 (1991);
W.~G.~Schmidt, S.~Glutsch, P.~H.~Hahn, and F.~Bechstedt,
%Efﬁcient O(N**2) method to solve the Bethe-Salpeter equation
Phys. Rev. B \textbf{67}, 085307 (2003).

\bibitem{Brandbyge2002}
M.~Brandbyge, J.~L.~Mozos, P.~Ordej{\'o}n, J.~Taylor
and K. Stokbro, Phys. Rev. B \textbf{65}, 165401 (2002).

\end{thebibliography}
\end{document}